\documentclass[one column,showpacs,keywords,floatfix,11pt,nofootinbib]{revtex4-1}
\usepackage{amssymb,amsmath}
\usepackage{graphicx,float}
\usepackage{indentfirst}
\usepackage{color}
\newcommand{\lie}{\pounds}

\newcommand{\be}{\begin{equation}}
\newcommand{\ee}{\end{equation}}

\newcommand{\ba}{\begin{eqnarray}}
\newcommand{\ea}{\end{eqnarray}}
\newcommand{\beq}{\begin{eqnarray}}
\newcommand{\eeq}{\end{eqnarray}}
\begin{document}

\title{Regular Bulk Solutions in Brane-worlds with Inhomogeneous Dust and Generalized Dark Radiation}
\author{A. Herrera-Aguilar}
\email{aha@fis.unam.mx}
\affiliation{Departamento de F\'{\i}sica, Universidad Aut\'onoma Metropolitana Iztapalapa,
San Rafael Atlixco 186, CP 09340, M\'exico D. F., M\'exico.}
\affiliation{Instituto de F\'{\i}sica y Matem\'{a}ticas, Universidad Michoacana de San Nicol\'as
de Hidalgo, Ciudad Universitaria, CP 58040, Morelia, Michoac\'{a}n, M\'{e}xico}
\author{A. M. Kuerten}
\email{andre.kuerten@ufabc.edu.br} \affiliation{CCNH, Universidade
Federal do ABC 09210-580, Santo Andr\'e, SP, Brazil}
\affiliation{Instituto de Ciencias F\'{\i}sicas, Universidad
Nacional Aut\'onoma de M\'exico
MEX--62210, Cuernavaca, Morelos, M\'{e}xico.}
\author{Rold\~ao da Rocha}
\email{roldao.rocha@ufabc.edu.br}
\affiliation{CMCC, Universidade Federal do ABC, 09210-580, Santo Andr\'e, SP, Brazil}
\affiliation{International School for Advanced Studies (SISSA), Via Bonomea 265, 34136 Trieste, Italy}
\pacs{04.50.Gh, 04.50.-h}

\begin{abstract}
From the dynamics of a brane-world with matter fields present in the
bulk, the bulk metric and the black string solution near the brane are
generalized, when both the dynamics of inhomogeneous dust/generalized dark radiation on the brane-world and inhomogeneous dark
radiation  in the bulk as well are considered --- as exact dynamical collapse solutions.
Based on the analysis on the inhomogeneous static exterior of a collapsing sphere of
homogeneous dark radiation on the brane, the associated black string warped horizon is studied,  as well as the 5D bulk metric near the brane. Moreover, the  black string and the bulk are  shown to be more regular upon time evolution, for suitable values for the dark radiation parameter in
the model, by analyzing the physical soft  singularities.
\end{abstract}

\maketitle

\affiliation{Departamento de F\'{\i}sica, Universidad Aut\'onoma Metropolitana Iztapalapa,
San Rafael Atlixco 186, CP 09340, M\'exico D. F., M\'exico} 
\affiliation{Instituto de F\'{\i}sica y Matem\'{a}ticas, Universidad Michoacana de San Nicol\'as
de Hidalgo, Ciudad Universitaria, CP 58040, Morelia, Michoac\'{a}n, M\'{e}xico}

\affiliation{Centro de
Ci\^encias Naturais e Humanas, Universidade
Federal do ABC 09210-580, Santo Andr\'e, SP, Brazil} 
\affiliation{(Instituto de Ciencias F\'{\i}sicas, Universidad
Nacional Aut\'onoma de M\'exico
MEX--62210, Cuernavaca, Morelos, M\'{e}xico.)}

\affiliation{Centro de Matem\'atica, Computa\c c\~ao e Cogni\c c\~ao,
Universidade Federal do ABC, 09210-580, Santo Andr\'e, SP, Brazil.}
\affiliation{International School for Advanced Studies (SISSA), Via Bonomea 265, 34136 Trieste, Italy.}

\flushbottom
%
%

\section{Introduction}

Brane-world models with a single extra dimension \cite{RS,RS1}  are decidedly a 5D phenomenological realization of Ho$%
\check{\mathrm{r}}$ava-Witten supergravity solutions~\cite{HW}, if  the moduli effects from compact extra dimensions can be ignored (for a review, see e. g. \cite%
{maartens}). {{}{The Ho$\check{\rm r}$ava-Witten solution \cite{HW} can be thought of as being effectively 5D, with an extra dimension that can be large, when collated to the fundamental scale. They provide the basis for the well-known Randall-Sundrum (RS) brane-world models \cite{RS,RS1}, that comprise the mirror symmetry and  a brane with tension as well, which counter balances the leverage of the negative bulk cosmological constant on the brane, encompassing furthermore the branes self-gravity \cite{maartens}.  In RS brane-world scenarios, our Universe is embedded in a 5D bulk of type AdS$_5$ \cite{RS1}.
The formalism to be used hereon employs a general metric for the brane-world, instead of the Minkowski metric in the standard RS model \cite{RS1}.}}

{{}{Brane-world black holes were comprehensively studied  
 in Randall-Sundrum like brane-world cosmologies \cite{gog2,harko1,maar2000}, where the dynamical equations on the brane are different from the general relativity ones. In fact, the brane-world  framework presents terms that handle both the effects of the free gravitational field in the bulk and of the brane embedding in the bulk as well. The imprint  of the nonlocal gravitational field in the bulk on the brane provides a splitting  into anisotropic stress, flux, and nonlocal energy density, where this last determines the tidal acceleration out of the brane, possibly opposing the formation of singularities  \cite{maartens}. Unlike the nonlocal energy density and flux, the nonlocal anisotropic stress is not ascertained by any evolution equation on the brane.  In particular, isotropy of the cosmic microwave background make the existence of the FRW  background be under risk. Adiabatic density perturbations are furthermore coupled to perturbations in the bulk field, making an open  system on the brane \cite{maar2000}. }}

{{}{Bulk effects in the cosmological dynamics of brane-world scenarios have been studied \cite{harko3}, also  in the context of thick brane-worlds \cite{nucaa,barbosa,euclides,Jardim:2011gg,nosso1}, brane-worlds with variable tension \cite{hoff,bs1,al11,al12} and Kalb-Ramond fields \cite{daro}, as well as in the context of more general brane-worlds and  bulks \cite{cor2014,gggg,dicke,daRocha2,Gergely:2003pn}. Moreover,  consequences of the gravitational collapse were proposed in the context
of brane-world scenarios in, e. g., \cite%
{zuleta,ovalle2,da01,kofinas,Bruni:2001fd}.
In addition, dark matter  was investigated already in \cite{boemer1} as a bulk
effect on the brane.}

{{}{Black strings can be thought as being extended objects endowed with an event horizon, in low energy string theory  \cite{Horowitz:1991cd}. The bulk metric near the brane and the black
string warped horizon along the extra dimension are here reviewed, based on previous developments \cite{chamblin,maartens}. Originally, a Schwarzschild black hole on the brane-world was shown to be a black string in a higher dimensional spacetime, what leads to the usual astrophysical properties of black holes  to be recovered in this scenario \cite{chamblin}. In this prototypical context,  the Kretschmann curvature invariants diverge when  the black string event horizon is approached along the axis of the black string. Several generalizations provide attempts  to preclude 
singularities both in the bulk and on the brane-world as well. When variable brane tension
scenarios are taken into account, the brane tension can gauge the Kretschmann scalars involved. For instance, regular bulk solutions and black strings were obtained in Friedmann-Robertson-Walker  brane-worlds under the E\"otv\"os law \cite{Bazeia:2014tua}, where the singularities related to the McVittie metric 
can be partially controlled as the cosmological time elapses. Indeed, for this type of metric the 5D physical soft  singularities in the bulk are alleviated as time elapses, providing a regular 5D bulk solution, as the 5D Kretschmann invariants do not diverge.
When other metrics are taken into account, for instance the Casadio-Fabbri-Mazzacurati metric one \cite{casadiofm},  black strings can be still emulated \cite{daRocha:2013ki}, however the related singularities in the bulk remain, regardless.}}
 In order to accomplish it, effective/perturbative  approaches are usually employed, where the black string is made to evolve from the brane-world
\cite{chamblin,casadio1,maartens}.}

{{}{Brane solutions of static black hole exteriors with 5D corrections to the Schwarzschild metric have been found, for instance, in \cite{dadhich, starsgm, Visser}, and furthermore in the context where the  bulk singularities can be removed \cite{tamvakis}. }}
{{}{The (Schwarzchild) black string is unstable near the AdS$_5$ horizon, defining the so called Gregory-Laflamme instability \cite{gregory,pretor1}.}}
 This scenario might be drastically altered by the inhomogeneous dust and the dark radiation.
In order to accomplish this effect, we use a procedure to calculate both the metric near the brane and  the 5D black string horizon  \cite{bs1}, uniquely from a  brane-world black hole metric and the associated Weyl tensor. Based on the knowledge of both  the Sasaki-Shiromizu-Maeda effective field equations on the brane and upon the 5D
Einstein and Bianchi equations \cite{maartens,GCGR,Gergely:2006hd,maar2000, casadio1},  both the bulk metric near the brane and in particular the black string
warped horizon can be designed,   by using a Taylor expansion along the extra dimension. Such
procedure provides information about all the bulk metric components \cite{bs1}. 
Indeed, the bulk
spacetime may be either given, by solving the full 5D equations or  
alternatively obtained by evolving the brane-world black hole metric off the brane, what encompasses the imprint from the bulk via the Weyl tensor.  
  Numerical methods have been employed to find
black hole solutions in the context of black strings and fluid/gravity
correspondence \cite{pretor1}. Similar methods involving expansions of the
metric have been used in the context of black strings \cite{Haddad:2012ss}, 
disposing the black string metric as the leading order
solution in a Taylor expansion. 

The
bulk shape of the black string horizon has  been merely investigated in very particular
cases~\cite{maartens,casadio1}, and latterly the standard black
string was studied in the context of  a brane-world with variable tension~%
\cite{bs1}. Moreover, realistic models that take into account a post-Newtonian
parameter on the Casadio-Fabbri-Mazzacurati black string~\cite{plb2013}, and
the black string in a Friedmann-Robertson-Walker E\"otv\"os brane-world~\cite{daRocha:2013ki}, 
also represent interesting applications. 

Recently regular black strings solutions associated to a dynamical  brane-world have been  obtained in the context of a variable brane tension \cite{Bazeia:2014tua}. The analysis of the 5D Kretschmann invariants makes us capable of attenuating the bulk physical singularities along some eras of the evolution of the Universe for the McVittie metric on an E\"otv\"os fluid brane-world. {{}This paper is devoted to encompass a framework with dark radiation and inhomogeneous dust. The 5D physical singularities in the bulk are shown to be
inherited from the 4D brane-world and no additional singularity appears in
the bulk, for some range of parameters in our model. Nevertheless, the bulk physical soft  singularities can be unexpectedly controlled in the bulk upon time evolution, what makes a regular bulk 5D solution in
most ranges of the dark radiation parameter.}

This paper is organized as follows: in Section II the dark radiation
dynamics on the brane is analyzed and reviewed. Starting with the Lema\^{\i}tre-Tolman-Bondi  (LTB)  metric on the brane, 
the effective field equations for dark radiation on the brane are solved.
The dynamical radiation model is shown to mimic a 4D cosmological constant
on the brane. Both the black string solution and the bulk metric are obtained thereon. After obtaining the standard dark radiation 
model, a generalized framework is proposed. Both associated metrics are derived. In
Section III the 5D bulk metric near the brane and the generalized black string are derived and studied, and in Section IV
the black string warped horizon in the context of inhomogeneous dust and
generalized dark radiation is studied. Moreover, the
black string physical singularities are analyzed from the Kretschmann
invariants. The 5D physical singularities in the bulk reflect the 4D
brane-world physical singularities. We analyze further the Kretschmann
scalars generated by higher order derivatives of the Riemann tensor, and the respective physical soft  singularities show that the bulk 5D solution is regular, in some ranges of the dark
radiation parameter.

In order to fix the notation hereupon  {$\mu,\nu
=0,1,2,3$} and {$M,N=0,1,2,3,5$}, and let $n$ be a time-like covector field
normal to the brane and $y$ the associated Gaussian coordinate. The
brane metric components $g_{\mu \nu } $ and the corresponding components of the bulk
metric $\check{g}_{\mu \nu } $ are related by $g_{\mu \nu }+n_{\mu }\,n_{\nu }=\check{g}_{\mu \nu
}$~\cite{maartens}. With these choices we can write $g_{55}=1$ and $g_{\mu 5}=0$, and thus the 5D bulk metric reads
\begin{equation}\label{bulkmetric11}\check{g}%
_{MN}\,dx^{M}\,dx^{N}=g_{\mu \nu }(x^{\alpha },y)\,dx^{\mu }\,dx^{\nu
}+dy^{2}\,,\end{equation} where  $M,N=0,1,2,3$ effectively.

{{}{ The initial paradigm concerning a perturbative method 
 for obtaining the black string solution consisting in assuming the 
Schwarzschild form for the induced brane metric, on a RS brane-world.  Subsequently 
a sheaf of such solutions are disposed into the extra dimension~\cite{chamblin}:
 \beq
{}^{(5)} ds^2 &=& e^{-2|y|/\ell}\left(-\left(1-{\frac{2GM}{r}}\right)dt^2+ {dr^2\over 1-\frac{2GM}{r}}+r^2d \Omega_2\right) +
dy^2\, \label{bsss1}\,,
 \eeq\noindent where $\ell=\sqrt{\frac{-6}{\Lambda_5}}$ denotes the curvature radius of the bulk AdS$_5$, wherein the RS brane-world is embedded. 
Each space of constant $y$ is a 4D Schwarzschild
spacetime with a singularity along $r=0$ for all
$y$, the well-known (Schwarzschild) black string.

As it is going to be clearer in Section III, the areal radius of the sheaf of such solutions along the extra dimension is called the black string warp horizon, that shall be precisely defined in Section III. 
}}

\section{Dark Radiation Dynamics on the Brane}

\label{bulkmetric} Henceforward some results 
concerning the dark radiation dynamics on the brane will be shortly
revisited \cite{NVA,NV,Neves1}, in order to briefly introduce the framework to get  both the bulk metric near the brane and the black string 
encompassing the dark radiation parameter and the effective cosmological
constant. New black strings solutions are here derived in the scenario provided in \cite{NVA,NV,Neves1} on
a background described by a Lema\^{\i}tre-Tolman-Bondi (LTB) metric \cite%
{ltbl,ltbt,ltbb}. A solution for the black string that has as limit a tidal
Reissner-Nordstr\"{o}m black hole solution on the brane was obtained in \cite%
{NVA}, in the Randall-Sundrum scenario. In order to work with  the
effective Einstein equations on the brane, some conditions on the
projected Weyl tensor are usually assumed, in order  to provide a closed
system. Besides, by taking into account a system of equations where a specific state equation leads
to a inhomogeneous density that has precisely the dark radiation form  \cite{BDEL,Muko,bow} (and its generalizations for thick branes \cite{aqeel,nosso1}), and by solving the effective Einstein
equations, the LTB metric can be derived. Subsequently both the bulk near the brane and the black string
solution is obtained are obtained in this context.

From the dynamics of a brane-world with matter fields
present in the bulk \cite{NV}, the associated  black
string solution will be shown to present a generalized dark radiation form. The way to
obtain the LTB metric is essentially different from that acquired in \cite%
{NVA}. The inhomogeneous density is associated with conformal bulk fields, 
instead of being related to the electric part of the Weyl tensor. The black string solution is 
similarly obtained by a change in the coordinate system and its final form
generalizes the first case. 
Thereafter two new black string solutions will be presented and subsequently used
in the construction of the horizon profile in the bulk.

\subsection{The Lema\^itre-Tolman-Bondi Metric}

The Lema\^itre-Tolman-Bondi (LTB) metrics are exact solutions of the Einstein equations that  
describe inhomogeneous spacetimes, having dust as the source. This type of models consider inhomogeneous generalizations of the Friedmann-Robertson-Walker (FRW) metrics, as  the LTB metrics. An alternative approach for
the LTB space-time is based in evolution equations of covariant objects, as the density, expansion scalar, electric Weyl tensor, shear tensor and spatial
curvature. The dynamics is reduced to scalar equations, and the FRW
spacetime is achieved when two of these scalars associated with the shear
tensor and electric Weyl tensor are zero. This formulation is
based on a $1+3$ covariant description  \cite{Eve}, which can be further applied to  the LTB model \cite{Sussman, Sussman2}. In general, the applications of these models involve black
holes, galaxy clusters, superclusters, cosmic voids, supernovas and redshift
drift, for instance  \cite{BKHC}.
Initially found by Lema\^{\i}tre \cite{ltbl}, the LTB metric describes a
spherically symmetric inhomogeneous fluid with anisotropic pressure
cosmological constant are present for instance in the Tolman model \cite%
{ltbt}. To derive the LTB solution, in comoving coordinates 
the general form for the line element is given by:
\begin{equation}
ds_{(4)}^{2}={g}_{\mu \nu }dx^{\mu }dx^{\nu
}=-dt^{2}+A^2(r,t)dr^{2}+R^2(r,t)d\Omega _{2},  \label{altb}
\end{equation}%
where the 2-sphere area element is denoted by $d\Omega_2$, the energy-momentum tensor written as $T_{\nu}^{\;\mu }=diag(-\rho
,-\rho _{\Lambda },-\rho _{\Lambda },-\rho _{\Lambda })$, where $\rho$ denotes the energy density,  $\Lambda _{4}$ denotes the brane cosmological constant  with associated energy density 
$\rho
_{\Lambda }=\kappa _{4}^{-2}\Lambda _{4}$,  and $\kappa _{4}$ is the 4D
gravitational coupling constant. The Einstein equations, {{}for each one of the space diagonal components,} are given by the following expressions: 
\begin{eqnarray}
\label{eis1}
\!\!\!\!\!\!\!\!\!\frac{1}{R}\left\{ 2\partial _{t}^{2}R+\frac{1}{R}\left[
1+\left( \partial _{t}R\right) ^{2}-\left( \frac{\partial _{r}R}{A}\right)
^{2}\right] \right\} &=&\Lambda _{4},\\
\label{eis2}\frac{\partial _{t}^{2}R}{R}+\frac{%
\partial _{t}R}{R}\frac{\partial _{t}A}{A}+\frac{\partial _{r}R}{R}\frac{%
\partial _{r}A}{A^{3}}+\frac{\partial _{t}^{2}A}{A}-\frac{\partial _{r}^{2}R%
}{RA^{2}}
\partial _{t}\partial _{r}R &=&\Lambda _{4},\label{4443}\\
\frac{\partial _{t}A}{A}\partial _{r}R&=& \Lambda _{4}\,. 
  \label{4}
\end{eqnarray}%
The function $A(r,t)=g(r)\partial _{r}R$ satisfies  Eq.(\ref{4443}). By
setting $g(r)=\left( {1+f(r)}\right) ^{-1/2}$, the 
usual form of the LTB metric is hence obtained: 
\begin{equation}
ds_{(4)}^{2}=-dt^{2}+\frac{\left( \partial _{r}R\right) ^{2}}{1+f}%
dr^{2}+R^{2}d\Omega _{2},  \label{ltb}
\end{equation}%
where $f(r)>-1$. The function $f$ can be interpreted as the
energy density shell $f(r)=2E(r)$. The function $g(r)$ is a geometric factor such
that when $g(r)=1$ the spatial sections are flat. Eqs.(\ref{eis1}) and (\ref{eis2}) are not {}independent, leading to the expression
\begin{equation}
\left( \partial _{t}R\right) ^{2}=f+\frac{2M}{R}+\frac{\Lambda _{4}}{3}R^{2},
\label{ltb1}
\end{equation}%
where $M=M(r)$ is an arbitrary function of integration that gives the
gravitational mass within each comoving shell of coordinate radius $r$. 
{{}By definition of mass in \cite{ltbb}, one can write $2dm/dr=k_{4}^{2}\rho
AR^{2}$, which implies $M=\int d_{r}m\sqrt{1+f}dr$. The
non-relativistic limit $f\ll 1$ yields $M\sim m$, and Eq.(\ref{ltb1}) reads    
$\frac{1}{2}\left( \partial _{t}R\right) ^{2}-\frac{M}{R}=\frac{f}{2}$. The first term is interpreted as kinetic energy, the second stands for the 
Newtonian potential term and $f$ is twice the energy of the system, when $\Lambda_4 = 0$. Hence $M$ is the 
relativistic generalization of the Newtonian mass. When $f$ is
negligible, namely in the non-relativistic limit,  the spatial sections
are flat when $g=1$. The function $g$  provides the  energy in each spatial section, and thus carries the
information of curvature for each section.}

Finally, Eq. (\ref{ltb1}) implies that
\begin{equation}
t-t_{N}(r)=\int_{0}^{R}\left( {{f+\frac{2M}{R}+\frac{\Lambda _{4}R^{2}}{3}}}%
\right) ^{-1/2}\;dR,  \label{ltb3}
\end{equation}%
where $t_{N}$ is known as the \textquotedblleft bang time". 
Eq. (\ref{ltb1}) can
 be used to classify the LTB models into three classes. When $%
\Lambda _{4}=0$ it reads:
\begin{eqnarray}
-1<f<0&&\qquad \text{elliptic}\,,\nonumber\\
f=0&& \qquad \text{parabolic}\,,\nonumber\\
f>0&&\qquad \text{hyperbolic}\,.\nonumber\end{eqnarray}
 When $\Lambda _{4}\neq 0$ the potential $%
V(R) = \frac{2M}{R}+\frac{\Lambda _{4}}{3}R^{2}$ leads to a
different classification, depending on the sign of $\Lambda _{4}$.

\subsection{LTB Solution on the Brane}

In this Subsection the LTB solution associated with dark radiation on the
brane is reviewed, starting with the projected Einstein equations on the
brane in vacuum. Unlike the Reissner-Nordstr\"{o}m black hole, this new
solution has a specific  {dark radiation tidal charge} $Q$.
The 4D and 5D coupling constants are related by $\kappa _{4}^{2}=\frac{1%
}{6}\lambda \kappa _{5}^{4}$. The
field equations in 5D Einstein theory lead to the 
projected equations \cite{GCGR,GCGR1} 
\begin{equation}\label{gmunuu}
G_{\mu \nu }=-\Lambda _{4}g_{\mu \nu }+\kappa_4^2T_{\mu \nu }+\kappa
_{5}^{4}\Pi _{\mu \nu }-\mathcal{E}_{\mu \nu },
\end{equation}%
where $\Pi _{\mu \nu }$ is a term quadratic in the energy-momentum $T_{\mu
\nu }$ and {{}{provides high-energy corrections arising from the extrinsic curvature of the brane, what  increases the pressure and effective density of collapsing matter}. The term $\mathcal{E}_{\mu \nu }$ is the projection of the bulk Weyl tensor and provide Kaluza-Klein corrections originated from 5D graviton stresses \cite{Bruni:2001fd}, as the massive modes for the graviton in the linearized regime. For observers on the brane such stresses are nonlocal, in the sense that they are local density inhomogeneities on the brane generate Weyl curvature in the bulk backreacting nonlocally on the brane \cite{maar2000, Muko,bow, maartens,al11,al12}. }Therefore the vacuum equations are $G_{\mu \nu }=-\Lambda _{4}g_{\mu
\nu }-\mathcal{E}_{\mu \nu },$ where the Weyl projected tensor can be
 identified with a trace-free energy-momentum as $\mathcal{E}_{\mu
\nu }\sim \kappa _{4}^{2}T_{\mu \nu }$
, provided by \cite{RMar}%
\begin{equation}
\mathcal{E}_{\mu \nu }=\left( \frac{\kappa _{5}}{\kappa _{4}}\right) ^{4}%
\left[ U\left( v_{\mu }v_{\nu }+\frac{1}{3}h_{\mu \nu }\right) +P_{\mu \nu
}+2Q_{(\mu }v_{\nu )}\right] ,
\end{equation}%
where $U$ is the effective energy, $P_{\mu \nu }$ is an anisotropic stress
tensor, $Q_{\mu }$ is the effective energy flux, $v_{\mu }$ is a 4D velocity
vector satisfying $v_{\mu }v^{\mu }=-1$, and $h_{\mu \nu }$ is such that $%
v^{\mu }h_{\mu \nu }=0$, being thus possible to write  $h_{\mu \nu } = g_{\mu \nu } + v_\mu v_\nu$. 
A non-static spherically symmetric brane-world with $P_{\mu \nu }\neq 0$ can be described by the
line element 
\begin{equation}
ds_{(4)}^{2}=-{C^2(r,t)}dt^{2}+B^{2}(r,t)dr^{2}+R^{2}(r,t)d\Omega _{2}
\label{altbb1}
\end{equation}%
 with $Q_{\mu }=0$. The anisotropic stress tensor $P_{\mu \nu }$ can be
represented by 
$P_{\mu \nu }=P\left( r_{\mu }r_{\nu }+\frac{1}{3}h_{\mu \nu }\right)
$, 
where $P=P(r,t)$ is a scalar field and $r_{\mu }$ is the unit radial
vector. 
With these assumptions the electric part of the Weyl tensor yields 
\begin{equation}
\mathcal{E}_{\mu}^{\;\nu }=\left( \frac{\kappa _{5}}{\kappa _{4}}\right)
^{4}diag(\rho ,-p_{r},-p_{T},-p_{T}),
\end{equation}%
with $\rho =U$, $3p_{r}=U+2P$ and $3p_{T}=U-P$. 

By assuming the brane field equation  $\nabla
^{\mu }{\mathcal{E}}_{\mu \nu }=0$ \cite{maartens} and by considering the state equation $\rho =-p_{r}%
%
%
$, it yields  $\partial _{t}U+4\frac{\partial _{t}R}{R}U=0=\partial
_{r}U+4\frac{\partial _{r}R}{R}U$ \cite{NVA}, implying that 
\begin{equation}
U=\left( \frac{\kappa _{5}}{\kappa _{4}}\right) ^{4}\frac{Q}{R^{4}}\,,
\label{dnva} 
\end{equation}\noindent where the constant $Q$ is the dark radiation tidal charge \cite{NVA,NV,Neves1}.
As $\rho =U$, the energy density in this case  is related to the inhomogeneous density. Thus the 4D
Einstein equations (\ref{gmunuu}) read%
\begin{equation}
G_{\mu \nu }=-\Lambda _{4}g_{\mu \nu }-\frac{Q}{R^{4}}\left( v_{\mu }v_{\nu
}-2r_{\mu }r_{\nu }+h_{\mu \nu }\right) .
\end{equation}%
It follows, by solving the above equations, that  the component $G_{tr}$ 
is obtained by the expression $\partial _{t}B\partial _{r}R-B\partial
_{t}\partial _{r}R=0,$ and therefore the function \begin{equation}B=H^{-1}\partial _{r}R\label{hhh}\end{equation} satisfies this
relation with $H=H(r)$. Considering thus such expression for $B$ in the trace equation $%
-G_{t}^{\;t}+G_{r}^{\;r}+2G_{\theta }^{\;\theta}$   \cite{NVA} it is
possible to write the expression as follows 
\begin{equation}
\left( \partial _{t}R\right) ^{2}=f-\frac{Q}{R^{2}}+\frac{\Lambda _{4}}{3}%
R^{2},  \label{ltb11}
\end{equation}%
which is similar to Eq.(\ref{ltb1}). Integrating Eq. (\ref{ltb11}), it reads 
\begin{equation}
\pm t+\tau (r)=\int \left( {{f-\frac{Q}{R^{2}}+\frac{\Lambda _{4}R^{2}}{3}}}%
\right) ^{-1/2}\;dR,
\end{equation}%
which is analogous to Eq. (\ref{ltb3}). 
 It is thus possible to write (\ref{altbb1}) in the LTB form given by (\ref%
{ltb}). Making the transformation of the LTB coordinates ($t,r$) to
curvature coordinates ($T,R$) as 
\begin{equation}
T=t+\int \frac{R\sqrt{\frac{\Lambda _{4}}{3}-Q}}{\frac{\Lambda _{4}}{3}%
R^{4}-R^{2}-Q}\;dR,  \label{TR1}
\end{equation}%
the following 4D metric is finally obtained:  
\begin{equation}
ds_{(4)}^{2}=-\left( 1+\frac{Q}{R^{2}}-\frac{\Lambda _{4}}{3}R^{2}\right)
dT^{2}+\left( 1+\frac{Q}{R^{2}}-\frac{\Lambda _{4}}{3}R^{2}\right)
^{-1}dR+R^{2}d\Omega _{2}\,.  \label{tRN}
\end{equation}%
This metric is  known as the inhomogeneous static exterior of a collapsing sphere of
homogeneous dark radiation \cite{Bruni:2001fd,GoDa}. Note that when $\Lambda
_{4}=0$, the solution (\ref{tRN}) is formally analogous to the
Reissner-Nordstr\"{o}m solution, when one identifies the electric
charge to the dark radiation tidal charge. 

In what follows this solution will be 
generalized, by considering a generalized dark radiation term with  dark radiation charge $%
Q_{\eta }$ where $\eta $ is a parameter characterizing the model of the dark
radiation \cite{NV}. The dynamics of a spherically symmetric brane-world
is also analyzed, when the bulk a) carries matter fields; and b) when its warp
factor characterizes a global conformal transformation consistent with $%
\mathbb{Z}
_{2}$ symmetry. 
Finally, 
it is possible to study the bulk metric and the black string solutions with a  term analogous to
the black hole solution with cosmological constant on the brane. In this framework, the 
energy-momentum tensor encompasses conformal bulk matter fields, whose
dynamics provide a 
specific state equation \cite{NVA,NV,Neves1}.

Consider now a general conformal spherically symmetric metric  $d\mathring{s}_{5}^{2}$ consistent
with $%
\mathbb{Z}
_{2}$ symmetry along the extra dimension  on
the brane 
\begin{equation}\label{abr}
d\mathring{s}_{5}^{2}=\Omega ^{2}\left(-e^{2A}dt^{2}+e^{2B}dr^{2}+R^{2}
d\Omega_2+dz^{2}\right) ,
\end{equation}%
where $z$ stands for the conformal extra dimensional coordinate, and $A,B,R,$ and $\Omega$ are general functions of the coordinates $%
(t,r,z)$. $\Omega $ denotes the conformal factor.\ The Einstein field equations
are given by%
\begin{equation}
\mathring{G}_{MN}=-\kappa _{5}^{2}\left[ -\mathring{T}%
_{MN}+\lambda \delta (z-z_{0})\mathring{g}_{MN}+\Lambda
_{5}\delta _{MN}\right] ,\label{zz0}
\end{equation}%
where $\mathring{g}_{MN}$ denotes the components of the is the induced metric and $\mathring{T}_{\mu\nu}$ stands for  the components of the energy-momentum tensor representing the bulk
fields. The brane is localized at $z=z_{0}$. 

Under the conformal transformation $%
\mathring{T}{}^{MN}=\Omega ^{s}T{}^{MN}$, the energy-momentum tensor is
assumed, as usual, to have weight $s=-4$ \cite{NVA}. The conformal Einstein tensor is given by  
\begin{equation}
\mathring{G}_{MN}=G_{MN}+\Upsilon _{MN}(\check{g})\,.
\end{equation}%
By using 
the expression 
$\mathring{g}_{N}^{\;M}=\Omega ^{-1}g{}_{N}^{\;M}$, 
Eq.(\ref{zz0}) hence leads to the following equations:
\begin{eqnarray}
G_{N}^{\;M} &=&\kappa _{5}^{2}T{}_{N}^{\;M},  \label{NV1} \\
\Upsilon_{N}^{\;M} &=&-\kappa _{5}^{2}\Omega ^{2}\left[ \Omega
^{-1}\lambda \delta (z-z_{0})g_{N}^{\;M}+\Lambda _{5}\delta _{N}^{\;M}%
\right] .  \label{NV2}
\end{eqnarray}
Eq.(\ref{NV1}) evinces that the 5D Einstein tensor is related solely to the
presence of fields in the bulk, and is independent both on the brane tension $%
\lambda $ and  upon the bulk cosmological constant $\Lambda _{5}$. Moreover, Eq. (%
\ref{NV2}) provides the dynamics of the conformal factor. The divergence
condition leads to $\nabla _{M}T_{\;N}^{M}+\Omega ^{-1}\left[
3T_{N}^{\,M}\partial _{M}\Omega -T\partial _{N}\Omega \right] =0, $ and
from the Bianchi identity Eqs. (\ref{NV1}) and (\ref{NV2}) it implies that   
\begin{eqnarray}
\text{either}\qquad\qquad\qquad\qquad\qquad\qquad\nabla _{M}T_{\;N}^{M}&=&0 \label{NV31}\\
\text{or}\qquad\qquad\qquad\qquad\qquad\qquad3T_{N}^{\;M}\partial _{M}\Omega &=&T\partial _{N}\Omega \,.  \label{NV32}
\end{eqnarray}

\noindent Hence Eqs. (\ref{NV31}) and (\ref{NV32}) 
require that $2T_{z}^{\;z}=T_{\mu }^{\;\mu }$.
Now, by considering the energy-momentum tensor 
\begin{eqnarray}\label{ems1}
T_{N}^{\;M}=diag\left( -\rho
,p_{r},p_{T},p_{T},p_{z}\right)\,,\end{eqnarray} the state equation $\rho
-p_{r}-2p_{T}+2p_{z}=0 $ holds. 
It implies that 
$\nabla _{z}T_{z}^{\;z}=0$ and subsequently that  $ \partial _{z}p_{z}=0\,$
Thus $\rho $, $p_{r}$ and $p_{T}$ must be independent of $z$. 

The system of inhomogeneous dark radiation and an effective cosmological
constant is defined by a conformal bulk matter with the following equations
of state%
\begin{equation}
\rho =-p_{r}=\rho _{DR}+\frac{\Lambda }{\kappa _{5}^{2}}\text{ \ \ and \ \ }%
p_{T}+\eta \rho =\frac{\Lambda }{\kappa _{5}^{2}}\left( \eta -1\right) ,
\end{equation}%
where $\eta $ characterizes the dark radiation model and $\Lambda $ is a
bulk quantity and mimics a 4D cosmological constant on the brane. 
The components of the Einstein tensor can be thus evinced:
\begin{eqnarray}
G_{r}^{\;r} &=&G_{t}^{\;t}=-\kappa _{5}^{2}\rho _{DR}-\Lambda\,,\qquad\qquad
G{}_{\theta }^{\;\theta }=G_{\phi }^{\;\phi
}=-\kappa _{5}^{2}\eta \rho _{DR}-\Lambda\,,\\
G_{z}^{\;z} &=&-\kappa _{5}^{2}\rho _{DR}\left( 1+\eta \right) -2\Lambda\,.
\end{eqnarray}%
By taking the divergence of $T_{\nu}^{\;\mu }$, it reads%
\begin{equation}
\partial _{t}\rho _{DR}+2\rho _{DR}\frac{\partial _{t}R}{R}\left( 1-\eta
\right) =0=\partial _{r}\rho _{DR}+2\rho _{DR}\frac{\partial _{r}R}{R}\left(
1-\eta \right) ,
\end{equation}%
what consequently leads to 
\begin{equation}
\rho _{DR}=\frac{Q_{\eta }}{\kappa _{5}^{2}}R^{2\eta -2},  \label{NV4}
\end{equation}%
where here $Q_{\eta }=const$ is interpreted as a generalized dark radiation
tidal charge and behaves like (\ref{dnva}) when $\eta =-1$. Regarding  the
following energy conditions \cite{wald}%
\begin{eqnarray}
\rho &\geq &0\qquad\text{ and }\quad\rho +p_{i}\geq 0, \\
\rho +\sum_{i=1}^{n}p_{i} &\geq &0\qquad\text{ and }\quad\rho +p_{i}\geq 0, \\
\rho &\geq &\left\vert p_{i}\right\vert ,
\end{eqnarray}%
respectively known as weak, strong and dominant energy conditions,  for $\rho =0$ (by taking the weak energy condition) the equality $\rho
_{DR}=-\kappa _{5}^{-2}\Lambda $ holds, such that $\Lambda =-Q_{\eta }R^{2\eta
-2}\leq 0$ for $Q_{\eta }\geq 0$. 

One can realize that $\rho =-p_{r}$ regards
the weak condition and $\rho +p_{T}=(1-\eta )\rho _{DR}$ holds if $\eta \leq
1$, where here $\rho >0$ satisfies the first condition. By a similar
analysis in 5D we see that $\eta \leq 0$ (weak), and $\eta \leq 0$ (4D and
strong), $\eta \leq -1/3$ (5D and strong), $\left\vert \eta \right\vert \leq
1$ (4D and dominant) and $-2\leq \eta \leq 0$ (5D and dominant). If $Q_{\eta
}<0$, all the energy conditions are violated. Taking the trace equation and
then making two integrations, it yields%
\begin{equation}
(\partial _{t}R)^{2}=\frac{Q_{\eta }}{2\eta +1}R^{2\eta }+\frac{\Lambda }{3}%
R^{2}+f,\qquad\text{for $\eta \neq -1/2$\,,}
\end{equation}%
what generalizes Eq. (\ref{ltb11}), being   identical to
it when $\eta =-1$. By performing one more integration it reads 
\begin{equation}
\pm t+\tau =\int {\left[ \frac{Q_{\eta }}{2\eta +1}R^{2\eta }+\frac{\Lambda 
}{3}R^{2}+f\right] ^{-1/2}}\;dR,  \label{PR}
\end{equation}%
where $\pm t$ refers to expansion or collapse and $\tau$ corresponds to 
the evaluation of the function at $t=0$. The
condition $R(0,r)=r$ is taken on the hypersurface $t=0$.  The radial
equation  leads to $H=\sqrt{1+f}$ \cite{NV}, and thus the metric has
the 4D LTB form. Hence the 5D conformal line element reads 
\begin{equation}
d\mathring{s}_{(5)}^{2}=\Omega _{RS}^2\left( -dt^{2}+\left({\partial _{r}R}/{%
H}\right) ^{2}dr^{2}+R^{2}d\Omega _{2}+dz^{2}\right) ,
\end{equation}
where $\Omega _{RS}$ is the Randall-Sundrum warp factor \cite{RS1}. From the
dynamical dark radiation models the marginal bound $f=0$
 corresponds actually to static solutions. Finally, the transformation from LTB
coordinates $(t,r)$ to the curvature coordinates $(T,R)$  is given thus by
\begin{equation}
T=t+\int \frac{\sqrt{\frac{
Q_{\eta }}{2\eta +1}R^{2\eta }+\frac{\Lambda }{3}R^{2}}}{\frac{
Q_{\eta }}{2\eta +1}R^{2\eta }+\frac{\Lambda }{3}R^{2}-1}dR\,.  \label{tttt}
\end{equation}%
The black string solution is therefore obtained: 
\begin{equation}
ds_{(4)}^{2}=-\left( 1-\frac{Q_{\eta }}{2\eta +1}R^{2\eta }-\frac{\Lambda }{3%
}R^{2}\right) dT^{2}+\left( 1-\frac{Q_{\eta }}{2\eta +1}R^{2\eta }-\frac{%
\Lambda }{3}R^{2}\right) ^{-1}dR^{2}+R^{2}d\Omega _{2}.  \label{lbtt}
\end{equation}

The models that describe the inhomogeneous static exterior of a collapsing
sphere of homogeneous standard dark radiation require the value\textbf{\ $%
\eta =-1$} \cite{NVA,Bruni:2001fd,GoDa}. In
general the exterior spacetime is not static in the brane-world scenario  \cite{Bruni:2001fd}, however the
collapse of a homogeneous Kaluza-Klein energy density  is static, and can be identified to the dark radiation. Hence 
in the case \textbf{$\eta =-1$} the model of homogeneous dark radiation is
recovered. It is worth to emphasize that the exterior is static solely  when
the system  has tidal charge and cosmological constant, where the
 physical mass equals zero. When $%
\Lambda =0$, the zero mass limit of the tidal Reissner-Nordstr\"{o}m black
hole is obtained \cite{dadhich}. The event horizon is determined by the solutions of the following equation: 
\begin{equation}
1-\frac{Q_{\eta }}{2\eta +1}R^{2\eta }-\frac{\Lambda }{3}R^{2}=0
\label{noso}
\end{equation}%
on the brane. Thus for $\Lambda =0$ it implies that  $R_{h}^{2\eta }=(2\eta
+1)/Q_{\eta }$. When $\Lambda \neq 0$, the exact location of the horizons
can not be obtained, with exceptions for the values $\eta =-1$ and $\eta
=1/2$. Respectively, the horizons are given by:
\begin{eqnarray}
R_{h}^{(\pm )} &=&\left\{ \frac{3}{2\Lambda }\left[ 1\pm \left( 1+\frac{%
4Q_{-1}\Lambda }{3}\right) ^{1/2}\right] \right\} ^{1/2}\,,\quad\text{for $\eta =-1$},\label{hor1}
\\
\text{\ }R_{h}^{(\pm )} &=&\frac{3Q_{1/2}}{4\Lambda }\left[ -1\pm \left( 1+%
\frac{16\Lambda }{9Q_{1/2}^{2}}\right) ^{1/2}\right]\,,\quad\text{for $\eta =1/2$}\,.  \label{rhh}
\end{eqnarray}

The conditions $\Lambda <0$ and $Q_{1/2}>0$ imply two horizons, an inner and
an outer one. For two specific values of $Q_{1/2}$ the singularity is naked,
thus violating the cosmic censorship hypothesis \cite{penrose}. Yet, if $%
\Lambda >0$ and $Q_{1/2}\gtrless 0$, then there is a single horizon $R_{h}^{(\pm)}$%
.

\section{Bulk metric and the black string}

\label{gaussian} In this Section, the bulk metric near the brane as well as the black string  associated to a black hole on a brane-world are briefly introduced  \cite%
{maartens,bs1,GCGR}. {{}{Eq.(\ref{bsss1}) represents the black string metric. By denoting ${}^{(5)}\mathcal{R}_{\mu \nu \sigma \rho }$ the components of the 5D Riemann tensor, as the 5D Kretschmann invariant $
{}^{(5)} \mathcal{R}_{\alpha\beta\rho\sigma}\,{}^{(5)} \mathcal{R}^{\alpha\beta\rho\sigma} = {40\over \ell^4}+{48G^2M^2
\over r^6}\, e^{4|y|/\ell}$ is unbounded  as
$y\to\infty$~\cite{chamblin,maartens}, thus the Schwarzschild 
solution is not a good candidate neither for a brane-world black hole
nor for trying to remove at least some of the bulk singularities.
Hence a well-established perturbative method is employed to find 
both the bulk metric near the brane and, in particular, the black string 
warped horizon along the extra dimension. In what follows such framework 
is revisited. This shall be further  accomplished in Section IV in the context of inhomogeneous dust and generalized dark radiation on the brane-world and inhomogeneous dark
radiation  in the bulk as well. There we shall evince that in such scenario the bulk (and the black string) can be regular for certain ranges of the dark radiation parameters.}

In brane-worlds with $\mathbb{Z}_2$ symmetry, the junction
conditions imply that the extrinsic curvature of the
brane is given by \cite{maartens}}
\begin{equation}
K_{\mu \nu }=\frac{\kappa _{5}^{2}}{2}\left[- T_{\mu \nu }+\frac{1}{3}\left(T-
\lambda \right) g_{\mu \nu }\right] \ .  \label{curv}
\end{equation}%
The trace-free and symmetric  components of the bulk Weyl tensor $C_{\mu \nu
\sigma \rho }$ are respectively given by $\mathcal{B}_{\mu \nu \rho
}=g_{\mu }^{\;\tau }\,g_{\nu }^{\;\sigma }\,C_{\tau \sigma \rho \beta
}\,n^{\beta }$ and $\mathcal{E}_{\mu \nu }=C_{\mu \nu
\sigma \rho }\,n^{\sigma }\,n^{\rho }$, where $n^\alpha$  denote components of a vector field out of the braneworld. Hereupon we denote by $\mathcal{R}_{\mu \nu \sigma \rho }= {}^{(5)}\mathcal{R}_{\mu \nu \sigma \rho
}(x^{\alpha},0)$ the components of the 5D Riemann tensor computed on the
brane.

{{}The effective field equations encompass the 5D Bianchi and Einstein  equations \cite{GCGR,Gergely:2003pn,maartens}, which have been generalized in a variable brane tension framework as well  \cite{GERGELY2008,hoff,bs1,cor2014,Bazeia:2014tua}. The effective field equations are provided by the following system of equations:
\begin{eqnarray}
 {\bf \pounds}\, K_{\mu\nu}&=& -\frac{1}{6}\Lambda_5 g_{\mu\nu}- {\cal E}_{\mu\nu} +K_{\mu\rho}K^\rho
{}_\nu \label{expansionone}\\
{\bf \pounds}\, {\cal E}_{\mu\nu}  &=& \frac{1}{6}
\Lambda_5\!\left(K_{\mu\nu}\!-\!Kg_{\mu\nu}\right)\!+\!\nabla^\tau
{\cal B}_{\tau(\mu\nu)}
\!+\!{\cal R}_{\mu\rho\nu\sigma}K^{\rho\sigma} \!+\!K_{\rho[\mu}K_{\sigma]\nu}
K^{\rho\sigma}\!+\!3K^\rho{}_{(\mu}{\cal
E}_{\nu)\rho}-K{\cal E}_{\mu\nu} \label{expansiontwo}
\\  {\bf \pounds}\, {\cal B}_{\mu\nu\rho}&=&K_\rho{}^\sigma {\cal
B}_{\mu\nu\sigma}-
2\nabla_{[\mu}{\cal E}_{\nu]\rho} -2{\cal B}_{\rho\sigma [\mu }K_{\nu]}{}^\sigma
\label{expansionthree}
\\ {\bf \pounds}\, {\cal R}_{\mu\nu\rho\sigma}
&=&-2{\cal R}_{\mu\nu\tau [\rho}K_{\sigma]}{}^\tau
-\nabla_{\mu}{\cal B}_{[\rho\sigma]\nu}\,. \label{expansionfour}
\end{eqnarray}
They are worked out the boundary
condition $
 {\cal B}_{\mu\nu\rho} = 2\nabla_{[\mu}K_{\nu]\rho}$ on the brane \cite{maartens}.
{{}{These expressions can be employed to compute a  Taylor expansion of the brane-world metric along the extra dimension, what perturbatively defines the bulk metric near the brane and the black string horizon.} 
The set of equations obtained from the 5D Einstein and Bianchi equations by
Shiromizu, Sasaki and Maeda ~\cite{maartens,GCGR} provides the  bulk metric near the brane, given by the standard Taylor expansion along the extra dimension $y$:
\begin{eqnarray}
\hspace*{-0.3cm}g_{\mu\nu}(x^\alpha,y)&=& g_{\mu\nu}(x^\alpha,0) +\left({\bf \pounds}\,g_{\mu\nu}(x^\alpha,y)\right)\vert_{y=0}\,|y| +\left({\bf \pounds}\,\left({\bf \pounds}\,g_{\mu\nu}(x^\alpha,y)\right)\right)\vert_{y=0}\,\frac{|y|^2}{2!} \nonumber\\
&& + \cdots + (\lie^k (g_{\mu\nu}(x^\alpha,y))\vert_{y=0} \frac{|y|^k}{k!} + \cdots\label{liederivative}
\end{eqnarray}
As comprehensively studied in \cite{bs1}, the term in the first order $|y|$ above 
is obtained from the definition of the extrinsic curvature $K_{\mu\nu} = \frac{1}{2}\lie g_{\mu\nu}$ and from the junction condition (\ref{curv}) as well, by denoting $g_{\mu \nu }= g_{\mu \nu }(x^{\alpha},0)$ hereon.  The term in $y^2$ is proportional to $\lie K_{\mu\nu}$, at Eq.(\ref{expansionone}). Moreover, the term $K_{\mu\alpha}K^\alpha
{}_\nu$ in Eq.(\ref{expansionone}) can be computed by  (\ref{curv}). Besides, the coefficient term of $|y|^3$ in Eq.(\ref{liederivative})
can be expressed as
\begin{equation} 2{\bf \pounds}\,\left({\bf \pounds}\,K_{\mu\nu}\right)
=\left({\bf \pounds}\,\left(K_{\mu\alpha}K^\alpha
{}_\nu\right) - {\bf \pounds}\,{\cal E}_{\mu\nu}-\frac{\Lambda_5}{6}\lie g_{\mu\nu}\right)\,,\label{ult}\end{equation}\noindent in virtue of Eq.(\ref{expansionone}). As the Lie derivatives terms in the right hand side of this last expression are respectively given by Eq.(\ref{expansionone}), due to Eq.(\ref{expansiontwo}), and by the definition of the extrinsic curvature, one hence arrives at the expression for $|y|^3$ in Eq.(\ref{liederivative}). Finally, the term in  $y^4$ is acquired by a similar reasoning.}

One thus finds the above expansion up to the fourth order   is given by \cite{bs1}
\begin{eqnarray}
g_{\mu \nu }(x^{\alpha },y) &=&g_{\mu \nu }-\kappa _{5}^{2}\left[ \frac{1}{3}(\lambda -T)g_{\mu \nu }+T_{\mu \nu
}\right] \,|y|  \notag \\
&&+\left[ \frac{1}{2}\kappa _{5}^{4}\left(\frac{2}{3}(\lambda -T)T_{\mu \nu }+ T_{\mu \alpha }T^{\alpha }{}_{\nu
}\right) -2\mathcal{E}_{\mu \nu }-\frac{1%
}{3}\left(\Lambda _{5}- \frac{1}{6}\kappa _{5}^{4}(\lambda -T)^{2}\right)
g_{\mu \nu }\right] \,\frac{y^{2}}{2!}  \notag \\
&&+\left. \Bigg[2K_{\mu \beta }K_{\;\,\alpha }^{\beta }K_{\;\,\nu }^{\alpha
}-\mathcal{E}_{(\mu |\alpha }K_{\;\,|\nu )}^{\alpha }-\nabla ^{\rho }%
\mathcal{B}_{\rho (\mu \nu )}+\frac{1}{6}\Lambda _{5}g_{\mu \nu }K+K^{\alpha
\beta }\mathcal{R}_{\mu \alpha \nu \beta }-K\mathcal{E}_{\mu \nu }\right. 
\notag \\
&&\left. \quad \qquad \qquad +3K^{\alpha }{}_{(\mu }\mathcal{E}_{\nu )\alpha
}+K_{\mu \alpha }K_{\nu \beta }K^{\alpha \beta }-K^{2}K_{\mu \nu }\Bigg]\;%
\frac{|y|^{3}}{3!}\right.  \notag \\
&&+\left. \Bigg[\frac{\Lambda _{5}}{6}\left( \mathcal{R}-\frac{\Lambda _{5}}{%
3}+K^{2}\right) g_{\mu \nu }+\left( \frac{K^{2}}{3}-\Lambda _{5}\right)
K_{\mu \alpha }K_{\;\,\nu }^{\alpha }+(\mathcal{R}-\Lambda _{5}+2K^{2})%
\mathcal{E}_{\mu \nu }\right.  \notag \\
&&+\left. \left( K_{\;\,\tau }^{\alpha }K^{\tau \beta }+\mathcal{E}^{\alpha
\beta }+KK^{\alpha \beta }\right) \,\mathcal{R}_{\mu \alpha \nu \beta
}+K^{2}\,K\,K_{\mu \nu }-\frac{1}{6}\Lambda _{5}{\cal R}_{\mu \nu }+2K_{\mu \beta
}K_{\;\,\rho }^{\beta }K_{\;\,\alpha }^{\rho }K_{\;\,\nu }^{\alpha }\right. 
\notag \\
&&\left. +\frac{7}{2}KK_{\;\,\mu }^{\alpha }\mathcal{E}_{\nu \alpha }-\frac{7%
}{6}K^{\sigma \beta }K_{\mu }^{\;\,\alpha }\mathcal{R}_{\nu \sigma \alpha
\beta }+\mathcal{E}_{\mu \alpha }\left( \frac{1}{2}KK_{\;\,\nu }^{\alpha
}-3K_{\;\,\sigma }^{\alpha }K_{\;\,\nu }^{\sigma }\right) \right.  \notag \\
&&\left. -\frac{13}{2}K_{\mu \beta }\mathcal{E}_{\;\,\alpha }^{\beta
}K_{\;\,\nu }^{\alpha }-4K^{\alpha \beta }\mathcal{R}_{\mu \nu \gamma \alpha
}K_{\;\beta }^{\gamma }-K_{\mu \alpha }K_{\nu \beta }\mathcal{E}^{\alpha
\beta }\Bigg]\,\frac{y^{4}}{4!}+\cdots \,\right.  \label{taylore}
\end{eqnarray}%
where $H\equiv H^{\;\mu}_{\mu}$ and $H^{2}\equiv H_{\rho \sigma }H^{\rho
\sigma }$, for any tensor $H$ of rank two. 

The black string warped horizon \cite{clark} can be studied from  $g_{\theta \theta }(x^{\alpha },y)$ in Eq.~(%
\ref{taylore}). In fact, regarding a spherically symmetric 4D metric 
\begin{equation}
ds_{(4)}^{2}=-h_{1}(r)\,dt^{2}+h_{2}(r)\,dr^{2}+r^{2}\,d\Omega_2\,,
\end{equation}%
modelling  a brane black hole, the usual 4D areal radial coordinate $r$ is
related to the 5D metric in Eq.(\ref{taylore}) by the expression $r^2 = {g_{\theta \theta }(x^{\alpha },0)}$~%
\cite{bs1,casadio1}. The black string has a warped horizon on the
brane that has radius $\mathrm{r}=\sqrt{g_{\theta \theta }(x^{\alpha },0)}$%
, where $\mathrm{r}$ denotes the coordinate singularity 
which is solution of $g_{rr}^{-1}(\mathrm{r})=0$ (See Eq.~(II.6) of Ref.~\cite%
{Casadioharms}).  The radius of the black string
warped horizon is hence $%
\mathrm{r}(y)= \sqrt{g_{\theta \theta }(x^{\alpha },y)}$. The term $g_{\theta \theta }(x^{\alpha },y)$ in Eq.~(\ref{taylore}) for $\mu =\nu =\theta $  corresponds to the bulk (squared) areal radius, and includes both the black string horizon for $r=\mathrm{r}$, and in particular the brane black hole 
horizon for $y=0$.}

\section{Black Strings and Dark Dust}

In this Section we investigate the black string related to the
induced black hole on the brane given by Eq. (\ref{lbtt}). 
The black string warped horizon is well known to be the component $g_{\theta
\theta }(x^{\alpha },y)$ in Eq.(\ref{taylore}), 
evaluated at
event horizons in (\ref{rhh}). At first let us calculate
such component for an arbitrary $x^{\alpha }$. In order to accomplish it, 
the 4D energy-momentum tensor on the brane is given by the brane-world components in Eq.(\ref{ems1}), where 
\begin{eqnarray}
\rho =-p_{r} &=&\frac{Q_{\eta }}{\kappa _{5}^{2}}R^{2\eta -2}+\frac{\Lambda }{\kappa _{5}^{2}}\\p_{T}&=&-\eta \frac{Q_{\eta }}{\kappa
_{5}^{2}}R^{2\eta -2}-\frac{\Lambda }{\kappa _{5}^{2}},
\end{eqnarray}
and the expressions for the 
energy-momentum, the projected Weyl tensor components, the extrinsic curvature, and the Riemann tensor respectively given respectively by (\ref{a1a}-\ref{a4c}) in Appendix are thus  obtained. Hence the  component $_{\theta\theta}$ of the metric, corresponding to the black string horizon along the bulk (\ref{taylore}), can be
written as: 
\begin{eqnarray}  \label{taylor1}
\hspace*{0cm}g_{\theta \theta }(x^{\alpha },y)\, &=&R^{2}-\frac{R^{2}}{3}%
\left[ \lambda \kappa _{5}^{2}+Q_{\eta }R^{2\eta -2}\left( 4-\eta \right)
+3\Lambda \right] \,|y|  \notag \\
&&+\frac{R^{2}}{3}{\Big\{}Q_{\eta }R^{2\eta -2}\left( 1-\eta \right) +\frac{1%
}{6}\left( \lambda \kappa _{5}^{2}+2\left[ Q_{\eta }R^{2\eta -2}\left( \eta
+2\right) +3\Lambda \right] \right) ^{2}-\Lambda _{5}~  \notag \\
&&-\frac{1}{2}\left( \eta Q_{\eta }R^{2\eta -2}+\Lambda \right) \left[
2\lambda \kappa _{5}^{2}+Q_{\eta }R^{2\eta -2}\left( \eta +8\right)
+9\Lambda \right] {\Big \}}\,\frac{y^{2}}{2!}  \notag \\
&&+R^{2}{\bigg\{}\!\frac{1}{108}\left[ Q_{\eta }R^{2\eta -2}\!\left( \eta
\!-\!4\right) \!-\!\lambda \kappa _{5}^{2}\!-\!3\Lambda \right] ^{3}\!-\!%
\frac{Q_{\eta }}{18}R^{2\eta -2}\left( \eta -1\right) \left[ Q_{\eta
}R^{2\eta -2}\left( \eta -4\right) -\lambda \kappa _{5}^{2}-3\Lambda \right]
\notag \\
&&-\frac{1}{9}\Lambda _{5}\left[ Q_{\eta }R^{2\eta -2}\left( \eta -4\right)
-\lambda \kappa _{5}^{2}-3\Lambda \right] -\frac{Q_{\eta }}{18}R^{2\eta
-2}\left( 1-\eta \right) \left[ 2\lambda \kappa _{5}^{2}+Q_{\eta }R^{2\eta
-2}\left( \eta +2\right) \right]  \notag \\
&&+Q_{\eta }R^{2\eta -2}\left( \eta -1\right) \left[ \frac{Q_{\eta }}{2\eta
\!+\!1}R^{2\eta }\left( \eta -1\right) +1\right] \!+\!\frac{\kappa _{5}^{2}}{%
12}\alpha ^{2}\!\left( 1\!-\!\frac{Q_{\eta }}{2\eta +1}R^{2\eta }-\!\frac{%
\Lambda }{3}R^{2}\right) ^{-1}\!\!\exp \!\left( -\alpha T\right)  \notag \\
&&+\frac{Q_{\eta }}{6}R^{2\left( \eta -2\right) }\left( 2\eta -2\right)
\left( \eta +2\right) \left( 2\eta -3\right) \left( 1-\frac{Q_{\eta }}{2\eta
+1}R^{2\eta }-\frac{\Lambda }{3}R^{2}\right)  \notag \\
&&+\frac{1}{12}\Lambda _{5}\left[ Q_{\eta }R^{2\eta -2}\eta -\lambda \kappa
_{5}^{2}-\Lambda \right] -\frac{1}{2}\left( \frac{\eta Q_{\eta }}{2\eta +1}%
R^{2\eta -2}+\frac{\Lambda }{3}\right) \left[ Q_{\eta }R^{2\eta -2}\eta
+\lambda \kappa _{5}^{2}+\Lambda \right]  \notag \\
&&+\frac{Q_{\eta }}{12}R^{2\eta -2}\left( 1-\eta \right) \left[ Q_{\eta
}R^{2\eta -2}\left( \eta -4\right) -\lambda \kappa _{5}^{2}-3\Lambda \right]
\notag \\
&&+\frac{1}{216}\left[ Q_{\eta }R^{2\eta -2}\left( \eta -4\right) -\lambda
\kappa _{5}^{2}-3\Lambda \right] \left[ Q_{\eta }R^{2\eta -2}\left( 9\eta
-5\right) +3\lambda \kappa _{5}^{2}-3\Lambda \right] {\bigg\}}\frac{|y|^{3}}{%
3!}+\cdots
\end{eqnarray}%
The last expression does not explicitly include the terms of order $y^{4}/4!$
as they are extensive and awkward, although we shall consider such terms in
our subsequent analysis.

Our first analysis consists in regarding the variable $T=T(r,t)$, which
appears solely as the term $\exp\! \left(-\alpha T\right)$ at the order $%
|y|^3/3! $ in Eq. (\ref{taylor1}). The graphic below in Fig. \ref{fig102-1} shows the dependence of
the component $g_{\theta\theta}(x^\alpha,y)$ on $T=T(r,t)$. 
\begin{figure}[h]
\includegraphics[width=13pc]{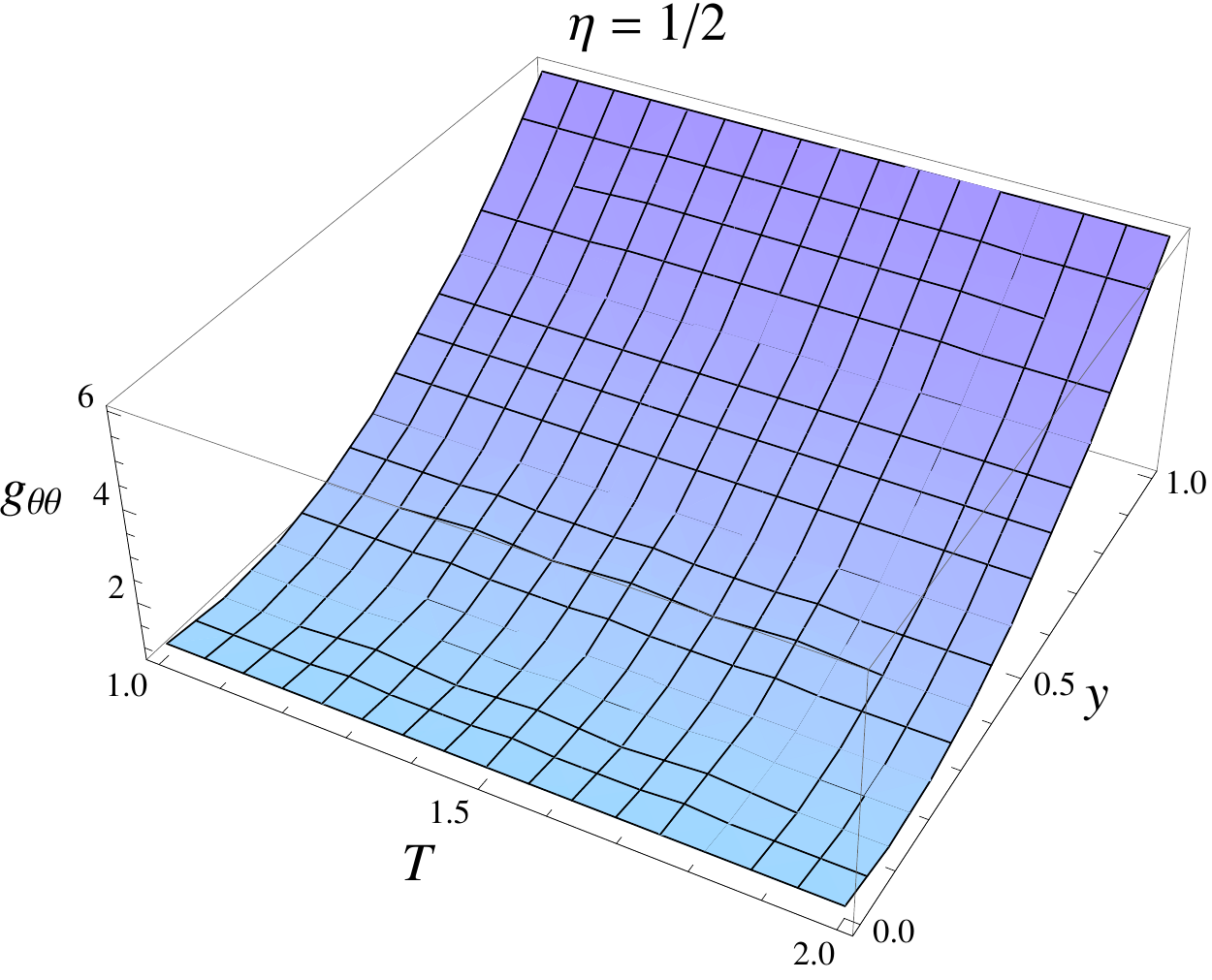}
\caption{{\protect\footnotesize Black string warped horizon $g_{\protect%
\theta\protect\theta}(R,y)$ along the extra dimension for $\Lambda = -1$,
for $\protect\alpha = 1$. }}
\label{fig102-1}
\end{figure}
Hence, near the brane, our results can be taken  to be independent of the
coordinate $T$. Below in Figs. \ref{fig1}, \ref{fig101}, and \ref{fig102} we depict the black string warped horizon $%
g_{\theta\theta}(R,y)$ as a function of the parameter $\eta$, for the
possible values $\Lambda=0,\pm1$: 
\begin{figure}[h]
\begin{minipage}{14pc}\includegraphics[width=2.1in]{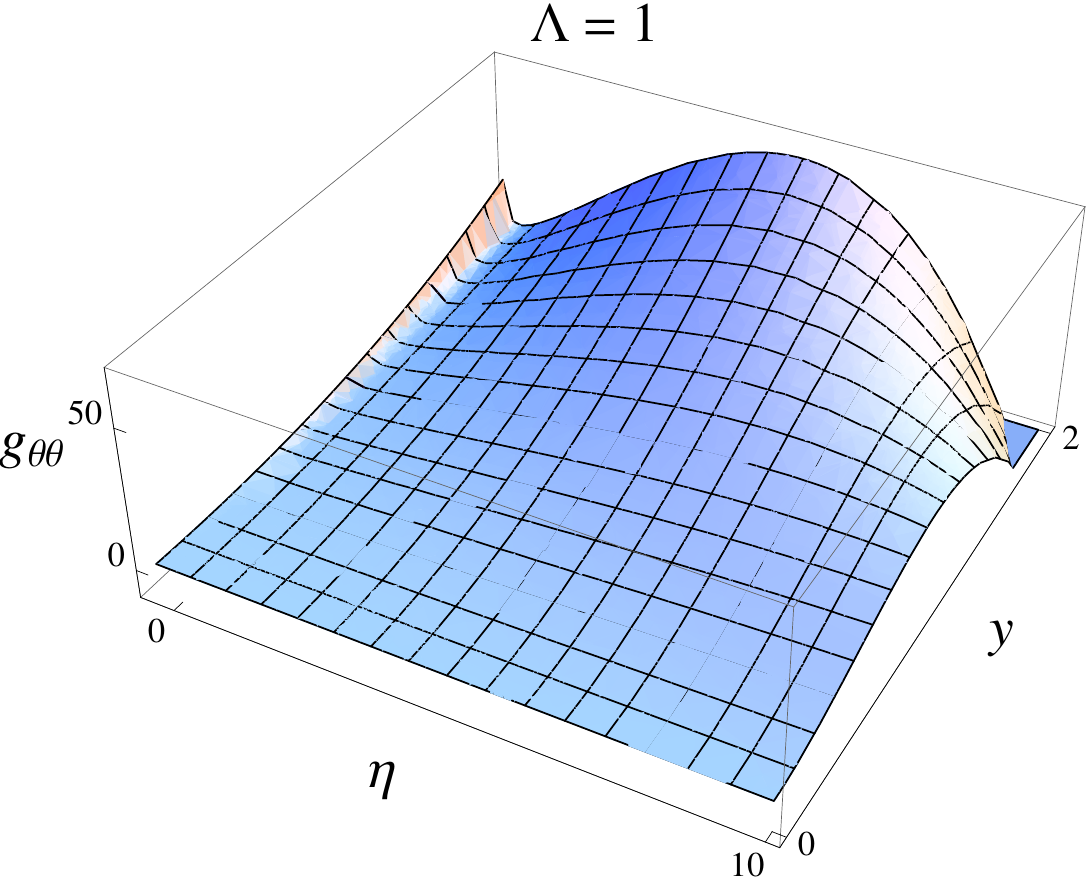}  
\caption{\footnotesize Black string warped horizon area  $g_{\theta\theta}(R,y)$ 
 along the extra dimension for $\Lambda = 1$ and the parameter $\eta$ variable.
 \label{fig1}}
\end{minipage}\hspace{7pc}%
\begin{minipage}{14pc}
\includegraphics[width=13pc]{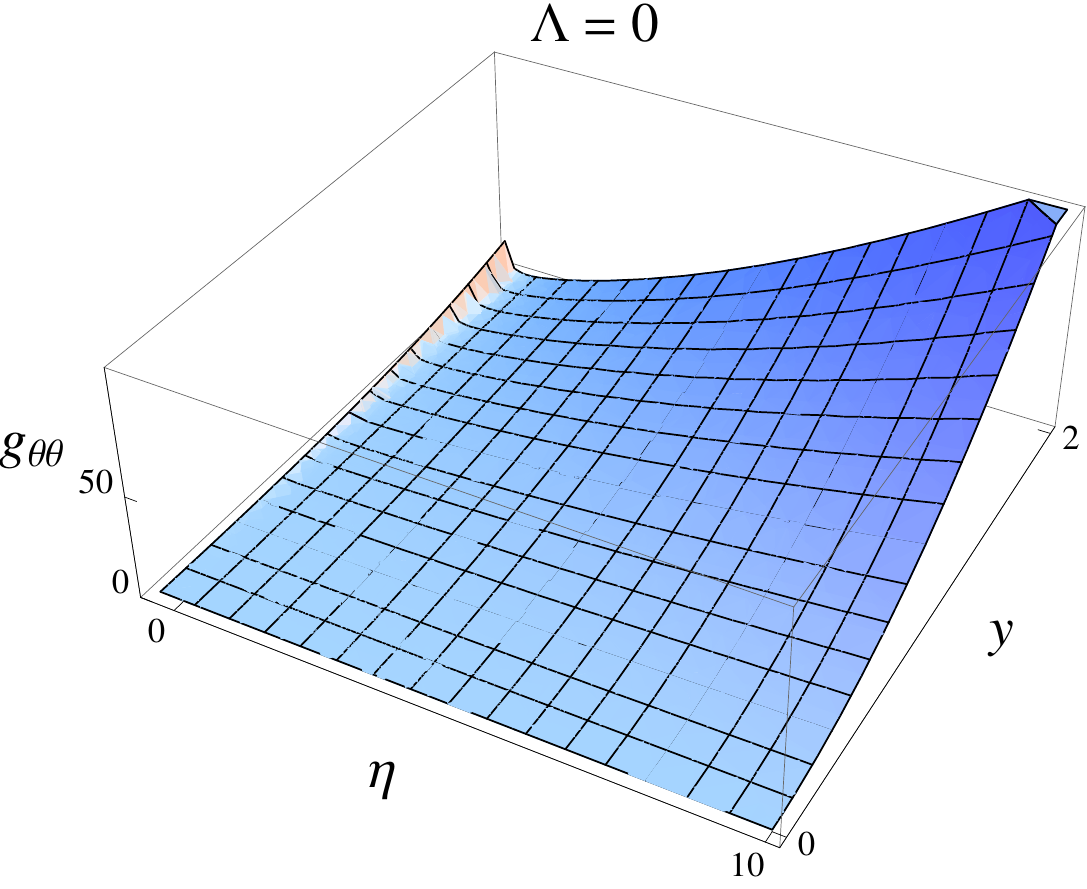}
\caption{\footnotesize Black string warped horizon area  $g_{\theta\theta}(R,y)$ 
 along the extra dimension for $\Lambda = 0$ and the parameter $\eta$ variable.
 \label{fig101}}
\end{minipage}
\end{figure}
\begin{figure}[h]
\includegraphics[width=13pc]{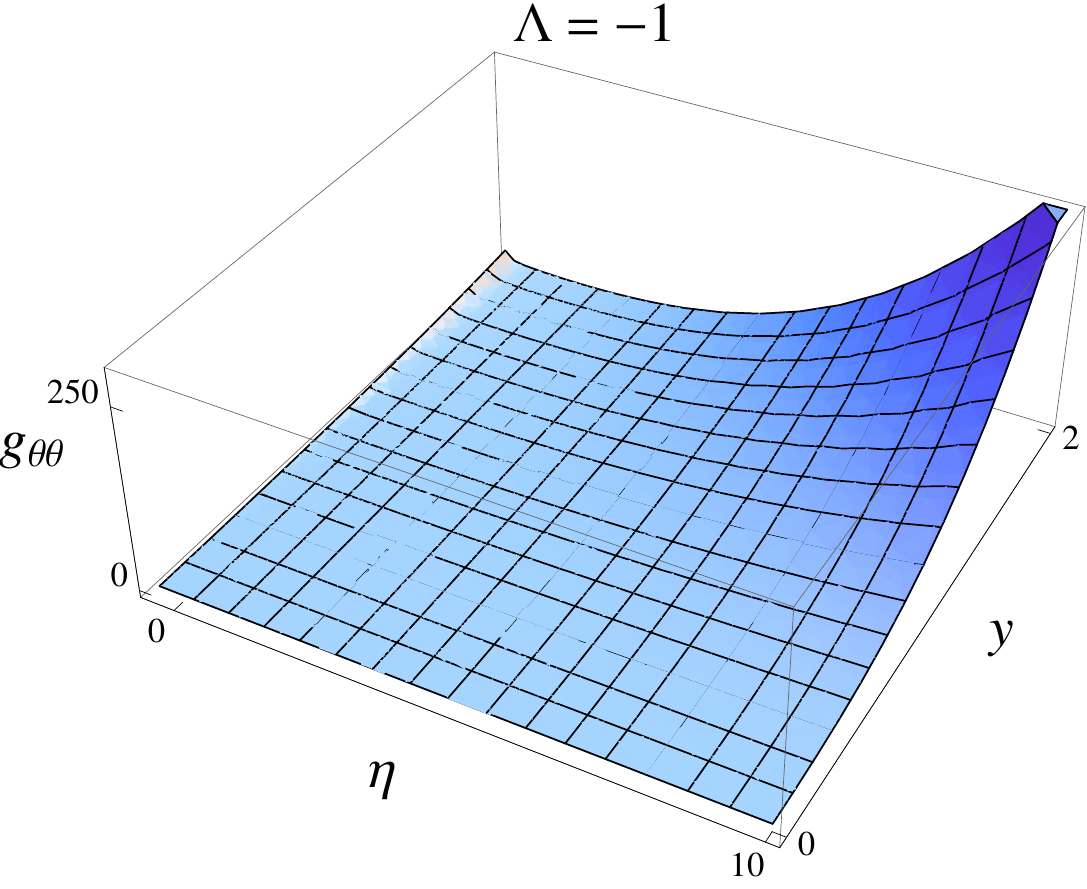}
\caption{{\protect\footnotesize Black string warped horizon area $g_{\protect%
\theta\protect\theta}(R,y)$ along the extra dimension for $\Lambda = -1$ and 
the parameter $\protect\eta$ variable. }}
\label{fig102}
\end{figure}\newpage
\noindent Fig. \ref{fig1} illustrates the case of a positive cosmological constant  $\Lambda=1$, where the dark radiation parameter $\eta$ plays an important role on the black string warped horizon profile. 
The bigger the value of $\eta$, the faster the warped horizon goes to zero along the extra dimension. In the range $1.8 \lesssim \eta\lesssim 8.7$ the black string warped horizon 
increases slower along the extra dimension, however it still goes to zero for higher values of the extra dimension. Figs. \ref{fig101} and \ref{fig102} show, respectively for $\Lambda = 0$ and $\Lambda =-1$, quite different profiles, as the black string warped horizon is shown monotonically  to increase   
along the extra dimension, as $\eta$ increases. 

Now, the black string warped horizon is analyzed for an
inhomogeneous static exterior of a collapsing sphere with homogeneous standard
dark radiation,  
as
a function of $Q_\eta$, as we have seen that when $\Lambda\neq 0$ in general it is not possible to obtain the exact location of the horizons by Eq.(\ref{noso}). In fact the two single exceptions are the models corresponding to $\eta = -1$ and $\eta = 1/2$, given by Eqs.(\ref{rhh}). 
\begin{figure}[h]
\begin{minipage}{14pc}
\includegraphics[width=16pc]{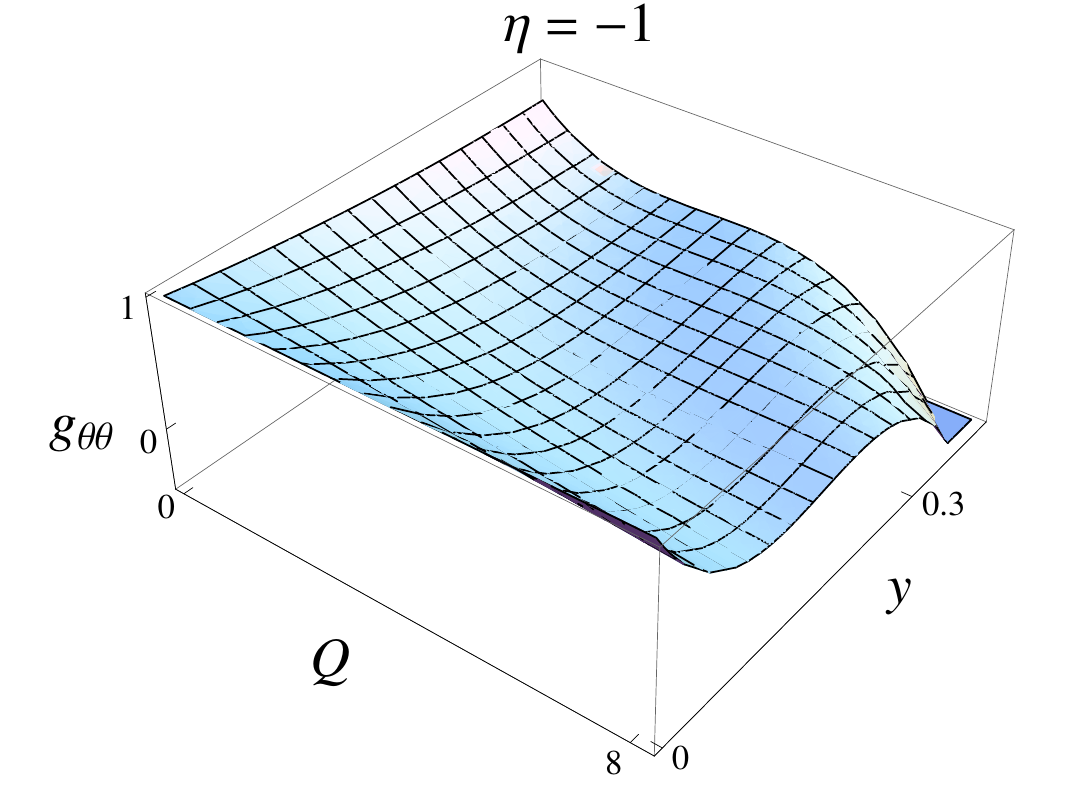}
\caption{\label{fig31} \footnotesize\; { Black string warped horizon area  $g_{\theta\theta}(R,y)$  along the extra dimension, as a function of the generalized dark radiation tidal charge parameter $Q_\eta$, for  $\eta=-1$.}}
\end{minipage}\hspace{7pc}%
\begin{minipage}{14pc}
\includegraphics[width=14pc]{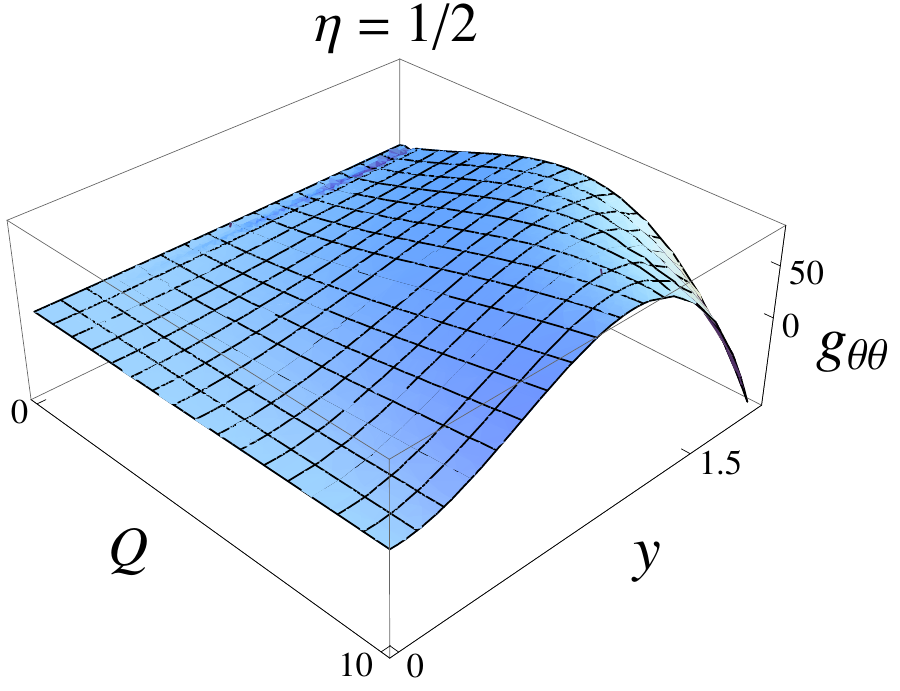}
\caption{\label{fig32} {\footnotesize\; Black string warped horizon area  $g_{\theta\theta}(R,y)$  along the extra dimension, as a function of $Q_\eta$ for  $\eta=1/2$.
}}\end{minipage}
\end{figure}
For fixed value of the dark radiation parameter, Figs. \ref{fig31} and \ref{fig32} show that the generalized dark radiation tidal charge $Q_\eta$ also influences the black string warped horizon along the extra dimension.
When $\eta = -1$, Fig. \ref{fig31} illustrates that the horizon goes to zero for some point along the extra dimension for $Q_{-1}\gtrsim5.1$, while Fig. \ref{fig32}  evinces that the horizon goes to zero for some point along the extra dimension for $Q_{1/2}\gtrsim1.2$.

Besides the black string warped horizon, Eq.(\ref{taylore}) provides more generally the bulk metric near the brane. In order to check whether the bulk is regular, we aim to study the 4D and the 5D Kretschmann invariants, related to the
black hole on the brane and to the black string in the bulk. Our analysis
of the bulk physical singularities is independent on the perturbative method
(\ref{taylore}), as the curvature invariants are  independent of it.

The Gauss equation which relates the 4D and 5D Riemann curvature tensors
according to 
\begin{equation}
{}^{(5)}\mathcal{R}^\mu _ {\;\; \nu\rho\sigma} = {}^{(4)}\mathcal{R}%
^\mu_{\;\; \nu\rho\sigma}  + K^\mu_ {\;\;
[\sigma}K_{|\nu|\rho]},  \label{gausseq}
\end{equation}
and the 4D and 5D Kretschmann invariants are given by 
\begin{eqnarray}  \label{4dk}
K^{(4)}&:=&\mathcal{R}^{\phi\psi\rho \sigma} \mathcal{R}_{\phi\psi\rho \sigma}
\\
K^{(5)}&:=& {}^{(5)}\mathcal{R}%
^{\phi\psi\rho \sigma}{}^{(5)}\mathcal{R}_{\phi\psi\rho \sigma}\,.  \label{5dk}
\end{eqnarray}
For the generalized dark radiation model they are related by: 
\begin{eqnarray}
K^{(5)} &=& K^{(4)} - R^2\left[ Q_{\eta }R^{2\eta -2}\left( \eta -4\right)
-\lambda \kappa _{5}^{2}-3\Lambda \right] \left[ \frac{\lambda k_{5}^{2}}{2}%
+Q_{\eta }R^{2\eta -2}\left( \eta -1\right) \right]\times  \notag \\
&&\!\!\!\!\!\!\left[\frac{2}{9}\!\left(1\!-\!\frac{Q_\eta R^{2\eta}}{2\eta+1}%
\!-\!\frac{\Lambda}{3}R^2\right)^{\!-2}\!\!\!\left( \frac{\eta Q_{\eta
}R^{2\eta -1}}{2\eta +1}\!+\!\frac{\Lambda }{3} R\right)+\frac{1}{36R^4}%
\left[ Q_{\eta }R^{2\eta -2}\left( \eta -4\right) -\lambda \kappa
_{5}^{2}-3\Lambda \right] \right]  \notag \\
&&+\frac{1}{18}\left(4\left[ \frac{1}{2}\lambda \kappa _{5}^{2}+Q_{\eta
}R^{2\eta -2}\left( \eta -1\right) \right] ^{2}+\left[ Q_{\eta }R^{2\eta
-2}\left( \eta -4\right) -\lambda \kappa _{5}^{2}-3\Lambda \right]
^{2}\right)^2  \label{kreto5}
\end{eqnarray}
The physical singularities are provided by $R = 0$ and by the solutions $R_h$ of the Eq. (\ref{noso}) for the values $\eta =-1$ and $\eta
=1/2$ given by Eqs.(\ref{rhh}).  
Neither new singularities are introduced nor the
existing ones ($R = 0$ and $R = R_h$ in Eqs.(\ref{hor1}) and (\ref{rhh}) are removed from the bulk.

We are going to show in what follows, by the analysis of the physical soft  singularities, that  the bulk can be regular. Indeed, the 4D invariant 
\begin{eqnarray}  \label{4dks}
\xi= (\nabla^\mu\nabla^\nu\mathcal{R}^{\phi\psi\rho \sigma})(\nabla_\mu\nabla_\nu\mathcal{R}_{\phi\psi\rho
\sigma})
\end{eqnarray}
(here $\nabla_\mu$ denotes the covariant derivative on the brane) is very
soft,  since it takes invariants involving at least two derivatives of the
curvature to detect it. Consequently, the 5D version of the invariant $\xi$
reads 
\begin{eqnarray}
{}^{(5)}\xi=(D^aD^b
{}^{(5)}\mathcal{R}^{\phi\kappa\zeta\sigma})(D_a D_b {}^{(5)}\mathcal{R}_{\phi\kappa\zeta \sigma}),  \label{xi5}
\end{eqnarray}
where $D_a$ denotes the 5D covariant derivative. It is worth to point
that $a,b$ are effectively 4D spacetime indexes, as the 5D covariant
derivative can be realized as $D_a = \nabla_\mu$ and $D_a = \nabla_5$, when the extra dimension $y$ is taken into account. It implies that the 5D Kretschmann invariant ${}^{(5)}\xi$ is given by 
\cite{plb2013} 
\begin{eqnarray}
^{(5)}\xi &=&\!\xi+ 2(\nabla_{\mu} \nabla_{\nu} K_{\tau[\rho\vert}K_{\psi%
\vert\sigma]})(\nabla^{\mu} \nabla^{\nu}K^{\tau\rho}K^{\psi\sigma}) -2
(\nabla_{y} \nabla_{\nu} K_{\tau\rho}K_{\psi\sigma}) (\nabla^{y}
\nabla^{\nu}K^{\tau\sigma}K^{\psi\rho})  \notag \\
&&+(\nabla_{(y} \nabla_{\nu)} R_{\tau\psi\rho \sigma})(\nabla^y \nabla^{\nu}
R^{\tau\psi\rho \sigma}) -2(\nabla_{(\mu} \nabla_{y)}
K_{\tau\rho}K_{\psi\sigma})(\nabla^{(\mu} \nabla^{y)} R^{\tau\psi\rho
\sigma})  \notag \\
&&- 4 (\nabla_{\mu} \nabla_{y} K_{\tau[\rho\vert}K_{\psi\vert\sigma]})
(\nabla^{\mu} \nabla^{y}K^{\tau\sigma}K^{\psi\rho})+ (\nabla_y^2
R_{\tau\psi\rho \sigma})((\nabla^y)^2 R^{\tau\psi\rho \sigma})  \notag \\
&&-4(\nabla_y^2 K_{\tau\rho}K_{\psi\sigma})((\nabla^y)^2 R^{\tau\psi\rho
\sigma})+2(\nabla_y^2 K_{\tau[\sigma\vert}K_{\psi\vert\rho]})((\nabla^y)^2
K^{\tau\sigma}K^{\psi\rho})\,. 
\end{eqnarray}%
\noindent Based on the values of the metric (\ref{lbtt}) and the
extrinsic curvature components given in (\ref{curv}), the 5D Kretschmann 
invariants $^{(5)}\xi$ for the bulk can calculated. 
Due
to the awkwardness of the expression for these invariants,  we opt to analyze our results by the
graphics in Figs. 8, 10, 12, and 14. 
\begin{figure}[ht]
\begin{minipage}{14pc}
\includegraphics[width=14pc]{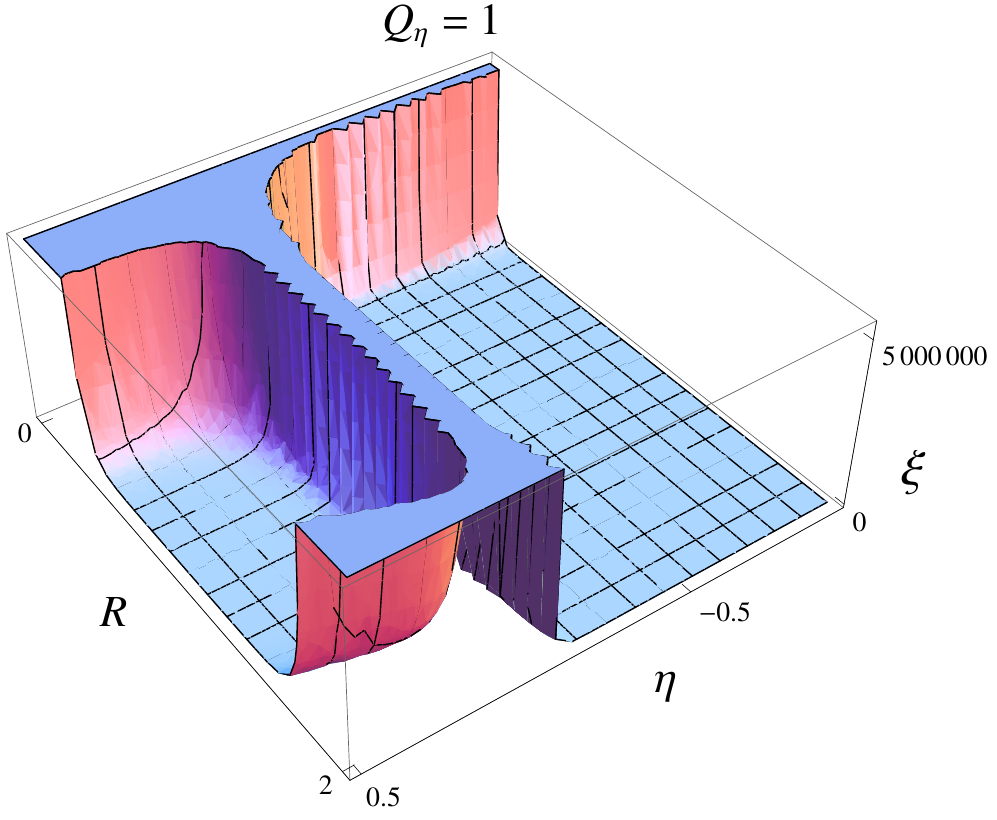}
\caption{\label{fig41} \footnotesize\; {4D Kretschmann invariant $\xi$ for $Q_\eta = 1$, for $\Lambda = 0$. 
}}
\end{minipage}\hspace{7pc}%
\begin{minipage}{14pc}
\includegraphics[width=15pc]{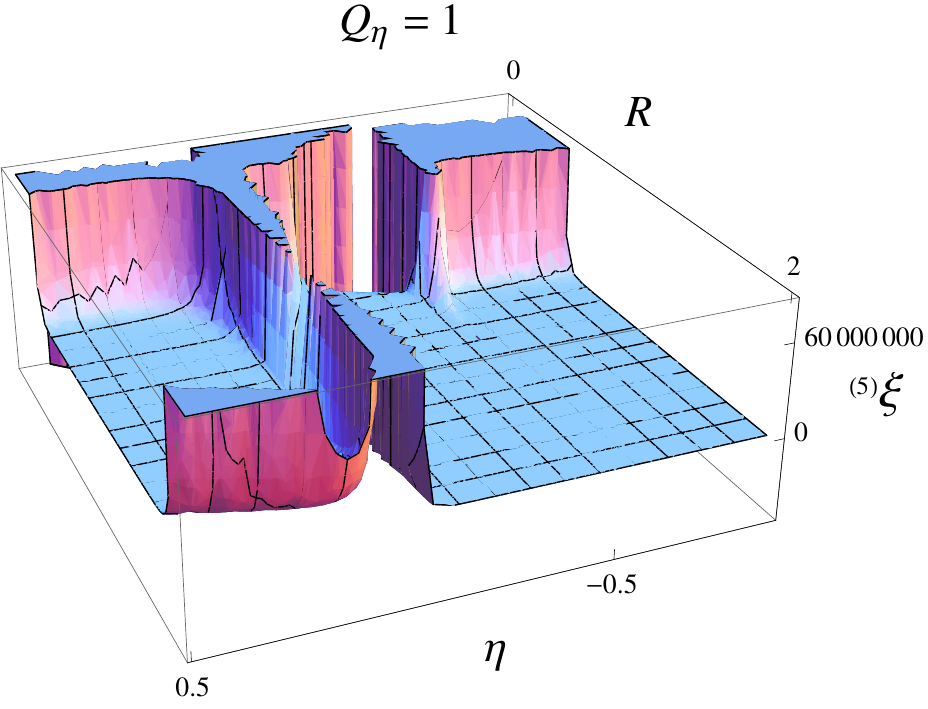}
\caption{\label{fig42} \footnotesize\; {5D Kretschmann invariant for $Q_\eta = 1$, for $\Lambda = 0$. 
}}
\end{minipage}
\end{figure}
Fig. \ref{fig41} has to be compared to Fig. \ref{fig42}, that describes the
5D invariant $^{(5)}\xi$. In Fig. \ref{fig41} since $\eta \sim
0$ then the 4D invariant $\xi$ diverges, independently of the value for $R$. On the other hand, Fig. \ref{fig42} depicts that for $\eta \sim 0$, the
5D Kretschmann scale ${}^{(5)}\xi$ goes to infinity for most values of $R$, but when 
$0.94\lesssim R\lesssim 1.06$ the 5D Kretschmann invariant $^{(5)}\xi$ does
not diverge. Hence for this range of physical soft  singularities present
on the brane the bulk is regular. Moreover, Fig. \ref{fig41} shows
that $\xi\rightarrow +\infty$ for $R\rightarrow 0$, however Fig. \ref{fig42}
evinces that for $0\lesssim \eta \lesssim 0.5$ the limit $\xi\rightarrow
+\infty$ on the brane alters to $^{(5)}\xi\rightarrow -\infty$, for values
of $R\sim  0$. Still, for $-1\lesssim \eta \lesssim -0.5$ the 5D
Kretschmann invariant $^{(5)}\xi$ does not diverge for $0.08\lesssim
R\lesssim 0.13$, and again this range of physical soft  singularities on the brane correspond to a regular bulk. Notwithstanding, now there
is a range $-1\lesssim \eta \lesssim -0.5$ and $0.08-0.13\lesssim R\lesssim
0.47-0.49$ that makes the 5D Kretschmann invariant $^{(5)}\xi$ to diverge.

Now the case $\Lambda = 1$ is regarded. 
\begin{figure}[ht]
\begin{minipage}{14pc}
\includegraphics[width=14pc]{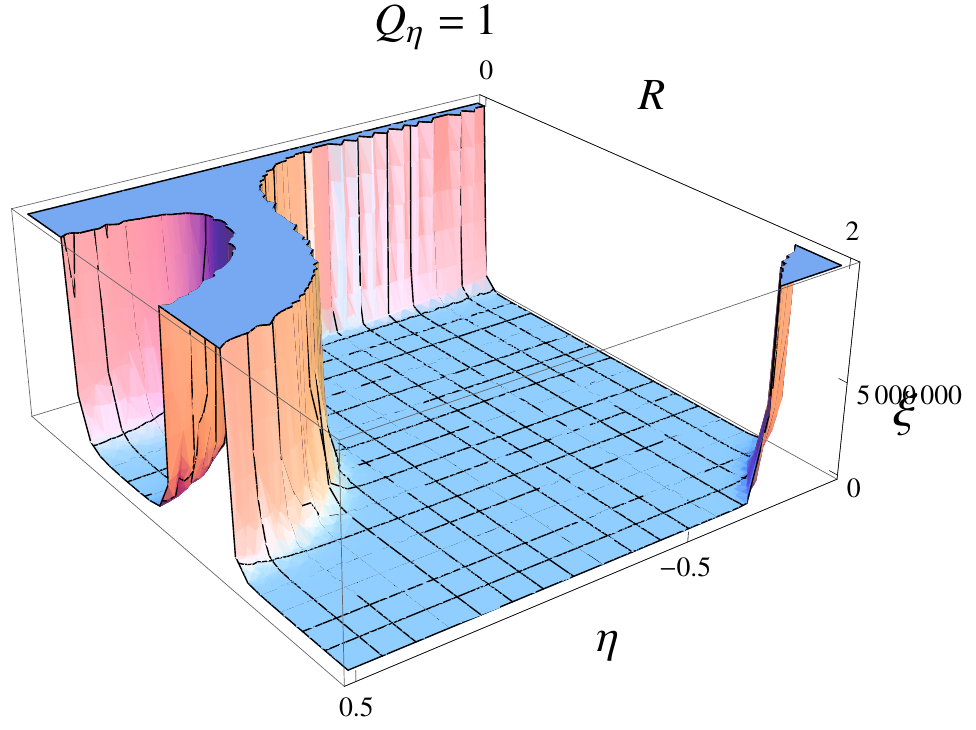}
\caption{\label{fig411} \footnotesize\; {4D Kretschmann invariant $\xi$ for $Q_\eta = 1$, for $\Lambda = 1$.}}
\end{minipage}\hspace{7pc}%
\begin{minipage}{14pc}
\includegraphics[width=15pc]{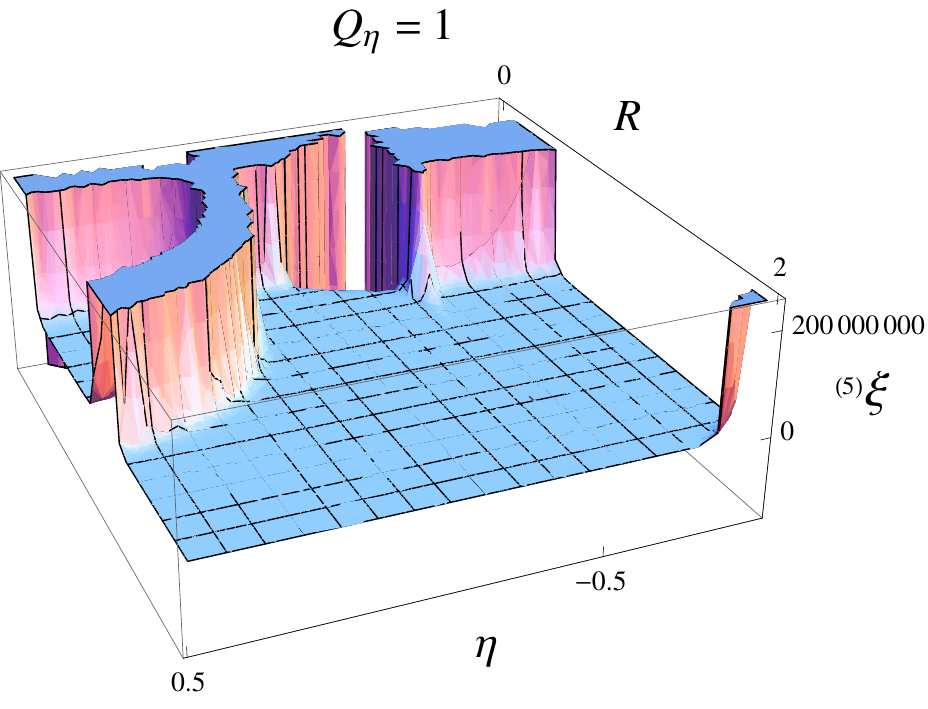}
\caption{\label{fig421} {5D Kretschmann invariant ${}^{(5)}\xi$ for $Q_\eta = 1$, for $\Lambda = 1$.
}}\end{minipage}
\hspace{7pc}
\end{figure}
Fig. \ref{fig411}, regarding the 4D invariant $\xi$, must be compared
respectively to Fig. \ref{fig421}, describing the 5D invariant $^{(5)}\xi$.
A similar pattern is realized in this case, where now Fig. \ref{fig411}
shows that $\xi\rightarrow +\infty$ for $R\rightarrow 0$.
We regard the $\Lambda=-1$ case in Figs. \ref{fig412} and \ref{fig422}, and the conclusions are analogous to 
both previous cases. 
\begin{figure}[ht]
\begin{minipage}{14pc}
\includegraphics[width=14pc]{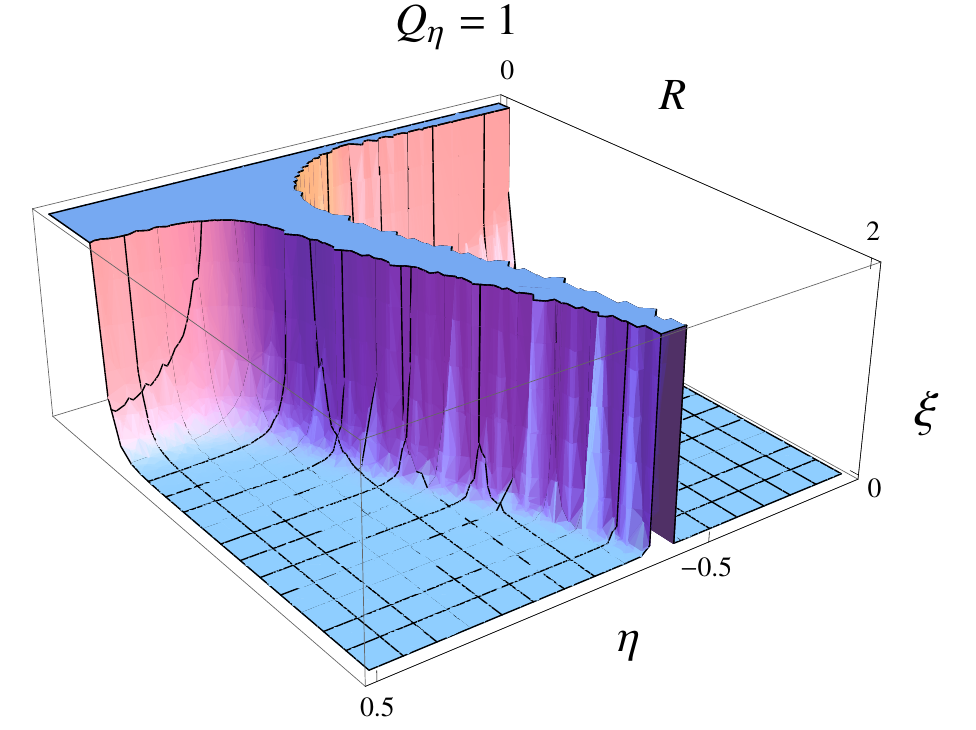}
\caption{\label{fig412} { 4D Kretschmann invariant $\xi$ for $Q_\eta = 1$, for $\Lambda = -1$. 
}}\end{minipage}\hspace{7pc}%
\begin{minipage}{14pc}
\includegraphics[width=15pc]{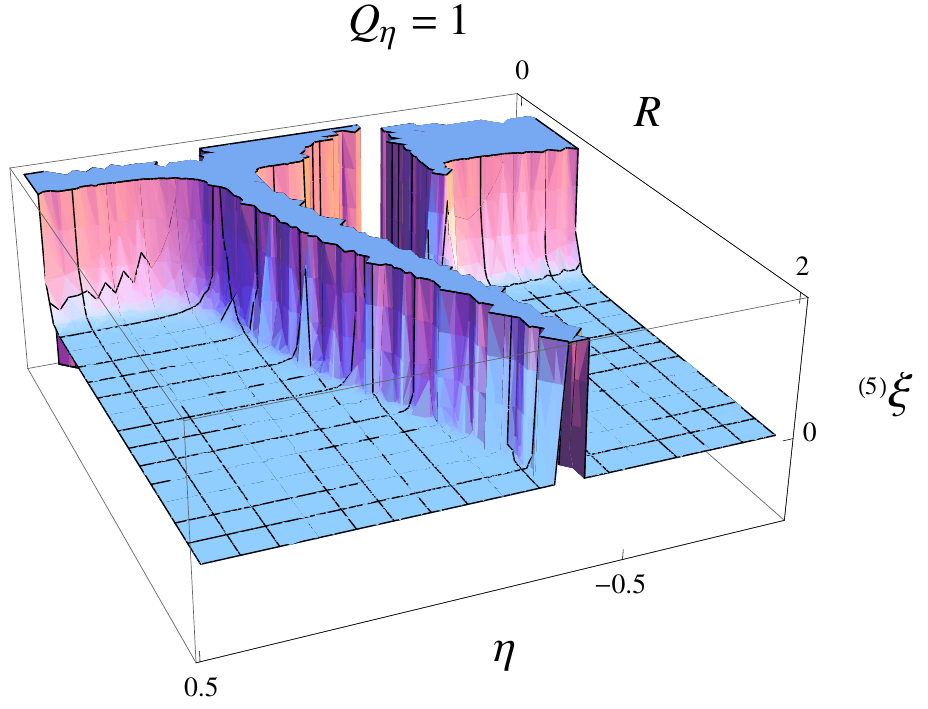}
\caption{\label{fig422} {5D Kretschmann invariant ${}^{(5)}\xi$ for $Q_\eta = 1$, for $\Lambda = -1$.
}}\end{minipage}
\end{figure}
Finally, the values of the 4D and the 5D Kretschmann invariants are provided
for $Q_\eta = 10$ and $\Lambda = 0$. The respective graphics for $\Lambda =
\pm1$ are quite similar.

\begin{figure}[ht]
\begin{minipage}{14pc}
\includegraphics[width=14pc]{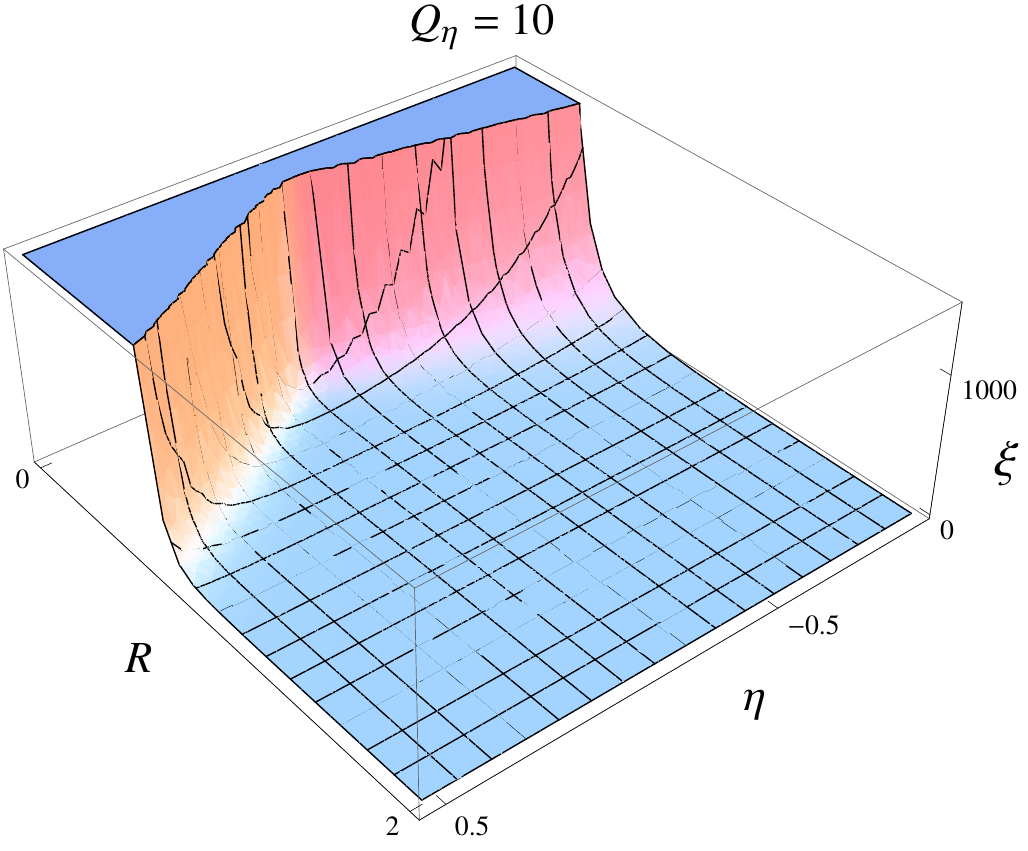}
\caption{\label{fig413} \footnotesize\; {4D Kretschmann invariant $\xi$ for $Q_\eta = 10$, for $\Lambda = 0$.
}}\end{minipage}\hspace{7pc}%
\begin{minipage}{15pc}
\includegraphics[width=14pc]{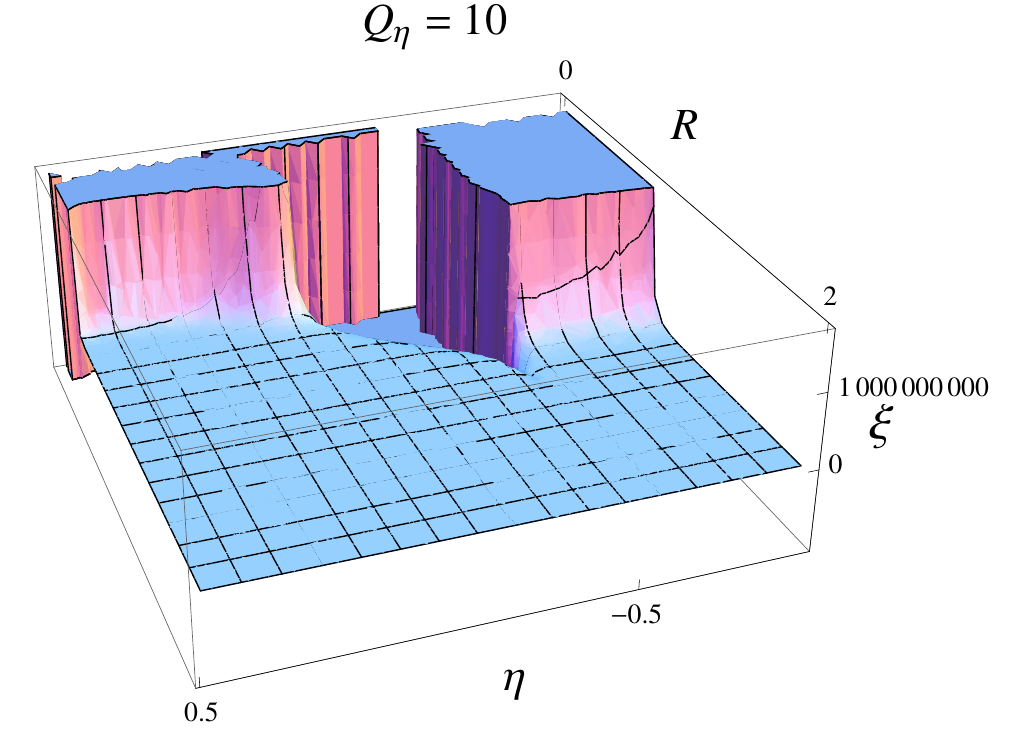}
\caption{\label{fig423} \footnotesize\; {5D Kretschmann invariant ${}^{(5)}\xi$ for $Q_\eta = 10$, for $\Lambda = 0$. }}
\end{minipage}
\end{figure}

Fig. \ref{fig413} illustrates that whatever the value for $\eta$ in the
range $[-1,1/2]$, the 5D curvature invariant $^{(5)}\xi$ is finite for $%
R\gtrsim 0.71$, and the whole bulk is regular, independent of the value of
the extra dimension $y$. Nevertheless, Fig. \ref{fig423} also evinces the
same  pattern of the previous cases. For $0\lesssim \eta \lesssim
0.5 $ the limit $\xi\rightarrow +\infty$ on the brane is modified into the
bulk to $^{(5)}\xi\rightarrow -\infty$, for values of $R$ near $R = 0$.
Still, again for $-1\lesssim \eta \lesssim -0.5$ the 5D Kretschmann
invariant $^{(5)}\xi$ does not diverge for $0.08\lesssim R\lesssim 0.13$,
and again this range of physical soft  singularities, present on the brane,
are banished in the bulk, as the Kretschmann invariants remain finite from a starting finite value for $R$.

\section{Concluding Remarks}

The bulk metric near the brane was obtained and in particular the black
string warped horizon, for  inhomogeneous dust and generalized dark radiation on the brane-world and inhomogeneous dark
radiation  in the bulk are regarded. The standard dark radiation \cite{NVA,BDEL,Muko,bow}
is a particular case analyzed, where the dark radiation parameter $\eta =-1$
and the mimicked cosmological constant $\Lambda$ equals zero, provided by the 5D pressure also equal to zero.

By analyzing the 4D and 5D standard Kretschmann invariants, respectively
defined by Eqs. (\ref{4dk}) and (\ref{5dk}), the Gauss equation is shown to
imply that the bulk associated to brane-world models with inhomogeneous dust
and generalized dark radiation inherits the brane-world physical
singularities at $R = 0$ and at $R=R_h$ is the solution of Eq. (\ref{noso}). Although the black string warped
horizon is obtained as a particular case of the  5D bulk metric near the brane via an effective  method, the analysis of the 5D bulk
singularities relies on an exact method provided by the Gauss equation (\ref%
{gausseq}). 

The 4D and 5D Kretschmann invariants,  respectively given by (\ref{4dks}) and (\ref{xi5}), show that although some physical soft  singularities exist on the brane, {{}{the  5D bulk can be regular, for some parameters of the dark radiation parameter, for various values of the generalized dark radiation tidal charge parameter $Q_\eta$ analyzed. Figs. 7-14 show these features for some values of $Q_\eta$ and all possible values of the mimicked cosmological constant on the brane. In fact Figs. 7, 9, 11, and 13 depict the 4D Kretschmann invariants, while Figs. 8, 10, 12, and 14 illustrate the 5D ones, for all values of the mimicked 4D cosmological constant on the brane $\Lambda=0,\pm1$. Such figures evince that the bulk is more regular in what concerns the respective soft singularities, for certain ranges of the generalized dark radiation tidal charge parameter analyzed. }}

\subsection*{Acknowledgements}

AMK is grateful to CAPES and \emph{Programa Ci\^encia sem
Fronteiras} (CsF) for financial support and to the ICF, UNAM for
hospitality. RdR thanks to SISSA for the hospitality and to CNPq grants No. 303027/2012-6 and No. 
473326/2013-2 for partial financial support. RdR is also \emph{Bolsista da CAPES Proc. 10942/13-0.} 
 AHA thanks the ICF, UNAM and
UAM-Iztapalapa for hospitality; he is grateful as well to PAPIIT-UNAM,
IN103413-3, \textit{Teor\'ias de Kaluza-Klein, inflaci\'on y perturbaciones
gravitacionales}, and SNI for financial support.

\appendix
\section{Energy-momentum, Weyl tensor and extrinsic curvature components}
The components of the energy-momentum and the Weyl tensor are given respectively by
\begin{subequations}
\begin{eqnarray}
T_{\theta \theta }&=&-\frac{R^{2}}{\kappa _{5}^{2}}\left( \eta Q_{\eta
}R^{2\eta -2}+\Lambda \right) ,\qquad \text{ \ \ \ \ }T_{\varphi \varphi }=-%
\frac{R^{2}}{\kappa _{5}^{2}}\left( \eta Q_{\eta }R^{2\eta -2}+\Lambda
\right) \sin ^{2}\theta\label{a1a} \\
T_{TT} &=&\frac{2}{\kappa _{5}^{2}}\left( 1-\frac{Q_{\eta }}{2\eta +1}%
R^{2\eta }-\frac{\Lambda }{3}R^{2}\right) \left( Q_{\eta }R^{2\eta
-2}+\Lambda \right) ,   \label{curv1} \\
T_{RR} &=&-\frac{2}{\kappa _{5}^{2}}\left( 1-\frac{Q_{\eta }}{2\eta +1}%
R^{2\eta }-\frac{\Lambda }{3}R^{2}\right) ^{-1}\left( Q_{\eta }R^{2\eta
-2}+\Lambda \right) .
\end{eqnarray}%
\end{subequations}
%
%
%
\begin{subequations}
\begin{eqnarray}
\mathcal{E}_{\theta \theta }&=&-\frac{Q_{\eta }}{6}\left( 1-\eta \right)
R^{2\eta },\qquad\qquad\qquad\mathcal{E}_{\varphi \varphi }=\mathcal{E}%
_{\theta \theta }\sin ^{2}\theta ,  \label{a2a} \\
\mathcal{E}_{TT} &=&-\frac{Q_{\eta }}{3}\left( 1-\eta \right) R^{2\eta
-2}\left( 1-\frac{Q_{\eta }}{2\eta +1}R^{2\eta }-\frac{\Lambda }{3}%
R^{2}\right) \label{a2b},  \\
\mathcal{E}_{RR} &=&\frac{Q_{\eta }}{3}\left( 1-\eta \right) R^{2\eta
-2}\left( 1-\frac{Q_{\eta }}{2\eta +1}R^{2\eta }-\frac{\Lambda }{3}%
R^{2}\right) ^{-1}.\label{a2c}
\end{eqnarray}
\end{subequations}
For the extrinsic curvature, by computing (\ref{curv}) we obtain 
\begin{subequations}
\begin{eqnarray}
K_{\theta \theta }&=&\frac{R^{2}}{6}\left[ Q_{\eta }R^{2\eta -2}\left( \eta
-4\right) -\lambda \kappa _{5}^{2}-3\Lambda \right] ,\;\;\;\;\;\;\;\;
\;\;\;\;\;\;\; K_{\varphi \varphi }=K_{\theta \theta }\sin ^{2}\theta , 
\label{a3a} \\
K_{TT}&=&\frac{1}{3}\left( 1-\frac{Q_{\eta }}{2\eta +1}R^{2\eta }-\frac{%
\Lambda }{3}R^{2}\right) \left[ \frac{1}{2}\lambda \kappa _{5}^{2}+Q_{\eta
}R^{2\eta -2}\left( \eta -1\right) \right] ,  \label{a3b} \\
K_{RR}&=&-\frac{1}{3}\left( 1-\frac{Q_{\eta }}{2\eta +1}R^{2\eta }-\frac{%
\Lambda }{3}R^{2}\right) ^{-1}\left[ \frac{1}{2}\lambda \kappa
_{5}^{2}+Q_{\eta }R^{2\eta -2}\left( \eta -1\right) \right] \,.\label{a3c}
\end{eqnarray}%
\end{subequations}
Finally, the
necessary components of the Riemann tensor are given by 
\begin{subequations}
\begin{eqnarray}
\mathcal{R}_{\theta T\theta T} &=&\left( \frac{Q_{\eta }}{2\eta +1}R^{2\eta
}+\frac{\Lambda }{3}R^{2}-1\right) \left( \frac{\eta Q_{\eta }}{2\eta +1}%
R^{2\eta }+\frac{\Lambda }{3}R^{2}\right)\,,\label{a4a} \\
\mathcal{R}_{\theta R\theta R} &=&\left( 1-\frac{Q_{\eta }}{2\eta +1}%
R^{2\eta }-\frac{\Lambda }{3}R^{2}\right) ^{-1}\left( \frac{\eta Q_{\eta }}{%
2\eta +1}R^{2\eta }+\frac{\Lambda }{3}R^{2}\right) , \label{a4b}\\
\mathcal{R}_{\theta \varphi \theta \varphi } &=&R^{2}\sin ^{2}\theta \left[ 
\frac{Q_{\eta }}{2\eta +1}R^{2\eta }+\frac{\Lambda }{3}R^{2}\right] .\label{a4c}
\end{eqnarray}
\end{subequations}

\end{document}